\documentclass[twocolumn,secnumarabic,amssymb, nobibnotes, superscriptaddress, nofootinbib,aps,prd]{revtex4-2}
\usepackage[colorlinks,citecolor=blue,linkcolor=blue,anchorcolor=blue,filecolor=blue, urlcolor=blue]{hyperref}
\usepackage{natbib}

\usepackage{amsmath}
\usepackage{adjustbox}
\usepackage{graphicx}
\usepackage{booktabs}       
\usepackage{amsfonts}      
\usepackage{nicefrac}       
\usepackage{microtype}
\usepackage{cleveref}
\usepackage{xcolor}
\usepackage{multirow}
\usepackage{makecell}
\usepackage{array}
\usepackage{orcidlink}
\usepackage{float}
\usepackage{lineno}
\usepackage{slashed}
\usepackage{bm}
\definecolor{dora}{RGB}{255,150,0}

\usepackage[normalem]{ulem}
\usepackage{caption}
\DeclareCaptionJustification{fulljustify}{\leftskip=0pt \rightskip=0pt \parfillskip=0pt plus 1fil}
\usepackage{tikz}
\usepackage{tikz-feynman}
\usepackage{caption}
\usepackage{subcaption}
\usepackage{placeins}

\begin{document}

\author{Adamu Issifu \orcidlink{0000-0002-2843-835X}} 
\email{ai@academico.ufpb.br}
\affiliation{CFisUC, Department of Physics, University of Coimbra, 3004-516 Coimbra, Portugal}
\affiliation{Departamento de F\'isica, Instituto Tecnol\'ogico de Aeron\'autica, DCTA, 12228-900, S\~ao Jos\'e dos Campos, SP, Brazil} 
\affiliation{Laborat\'orio de Computa\c c\~ao Cient\'ifica Avan\c cada e Modelamento (Lab-CCAM), Brazil}

\author{Orlando Oliveira \orcidlink{0000-0001-8266-3796}}
\email{orlando@uc.pt}
\affiliation{CFisUC, Department of Physics, University of Coimbra, 3004-516 Coimbra, Portugal}

\author{Tobias Frederico \orcidlink{0000-0002-5497-5490}} 
\email{tobias@ita.br}
\affiliation{Departamento de F\'isica, Instituto Tecnol\'ogico de Aeron\'autica, DCTA, 12228-900, S\~ao Jos\'e dos Campos, SP, Brazil}
\affiliation{Laborat\'orio de Computa\c c\~ao Cient\'ifica Avan\c cada e Modelamento (Lab-CCAM), Brazil}

\title{Effective field theory of scalar glueballs: Form factors and interaction radii}

\begin{abstract}
We develop a gauge-invariant effective field theory for scalar glueball interactions based on the Yang-Mills gluon condensate. Starting from a coherent-state formulation of the Yang-Mills vacuum, we derive an effective Lagrangian for the scalar glueball and its interactions with gluons. The resulting crossing-symmetric amputated four-gluon Green's function is projected onto the color-singlet $J^{PC}=0^{++}$ channel, yielding a normalized scalar glueball form factor that factorizes into universal kinematic structures and an intrinsic glueball function. The formalism predicts a parameter-free effective interaction radius for the ground-state scalar glueball with resonance $f_0(1710)$, $r_{\rm int}=\sqrt{6}/m_\phi=0.28$ fm, which is in excellent agreement with recent lattice Yang-Mills determinations of the mass-radius ($r=0.263(31)$ fm). The predicted momentum dependence of the normalized form factor is consistent with lattice Yang-Mills gravitational form factor data, with the $f_0(1710)$ candidate providing the best overall fit and $r_{\rm int}$. Inclusion of the first excited $0^{++}$ glueball produces modest corrections to the interaction radius ($r_{\rm int}^*=0.240$ fm) while yielding stable and kinematically robust modifications of the form factor. This framework provides a systematic bridge between effective field theory and first-principles lattice Yang-Mills calculations of scalar glueball dynamics.
\end{abstract}

\maketitle

\section{Introduction}

Glueballs are color-singlet bound states composed entirely of gluons and constitute one of the most distinctive nonperturbative predictions of quantum chromodynamics (QCD)~\cite{Mathieu:2008me, Giacosa:2021brl}. Their existence follows directly from the self-interaction of gluons in non-Abelian gauge theories. Extensive lattice Yang--Mills calculations have established a well-defined glueball spectrum, with the lightest state carrying quantum numbers $J^{PC}=0^{++}$ and a mass of approximately $1.7~\mathrm{GeV}$~\cite{Morningstar:1999rf, Chen:2005mg, Lucini:2004my, Li:2013pta, Huber:2021yfy} (for a review, see \cite{Vadacchino:2023vnc, Morningstar:2024vjk}). Hadron spectroscopy has also led to the identification of several experimental glueball candidates with different quantum numbers \cite{Klempt:2007cp, Crede:2008vw, Chen:2022asf, BESIII:2010gmv, BESIII:2019wkp, BESIII:2023wfi}. Identifying an observed hadron as a glueball or glueball-like state from spectroscopy remains an open question, owing in part to the possible mixing of glueball states with conventional hadrons carrying the same quantum numbers \cite{BESIII:2010gmv, Athenodorou:2023ntf}. As a result, deriving the internal structure and interactions of glueballs directly from the underlying nonperturbative Yang--Mills dynamics remains a significant theoretical challenge~\cite{Shifman:1978bx, Shifman:1978by, DiGiacomo:2000irz, Eichmann_2014, Eichmann:2015nra}.

Early effective descriptions of scalar glueballs were based on the QCD trace anomaly and scale-symmetry breaking, leading to dilaton-like effective Lagrangians that successfully capture several low-energy properties of the scalar sector~\cite{Migdal:1982jp, Gomm:1985ut, PhysRevD.29.1424}. More recently, lattice Yang--Mills calculations have provided direct information on scalar glueball gravitational form factors and mass-radii~\cite{Abbott:2025irb}, opening the possibility of confronting effective descriptions with first-principles nonperturbative calculations. At the same time, studies of the four-gluon vertex have identified a well-defined scalar color-singlet channel through appropriate tensor projections, providing a natural framework for investigating glueball exchange in gluonic scattering processes~\cite{Colaco:2024gmt, Oliveira:2025tfq}.

In this work, we develop a gauge-invariant effective field theory (EFT) for scalar glueball interactions derived directly from the nonperturbative Yang--Mills gluon condensate~\cite{Shifman:1978by, DiGiacomo:2000irz}. Using a coherent-state description of the Yang--Mills vacuum~\cite{Glauber:1963tx, Weinberg:1995mt}, we derive an effective glueball Lagrangian and the corresponding glueball--gluon interactions without phenomenologically introducing the scalar sector, in contrast to traditional dilaton-based effective theories~\cite{Migdal:1982jp, Gomm:1985ut, PhysRevD.29.1424}. The resulting effective theory is then used to construct the glueball-mediated amputated four-gluon Green's function, which is projected onto the color-singlet $J^{PC}=0^{++}$ channel using channel-resolved scalar projectors closely related to the Bose-symmetric tensor basis introduced in recent lattice studies of the four-gluon vertex~\cite{Colaco:2024gmt, Oliveira:2025tfq}. This projection naturally separates universal Lorentz--color kinematic structures from the intrinsic scalar glueball dynamics and motivates the definition of a reduced, normalized, crossing-symmetric scalar glueball form factor.

The proposed framework yields several quantitative predictions. We derive a parameter-free expression for the effective interaction radius \cite{Gao:2021sml, Sick:2018fzn, PhysRev.126.2256} of the ground-state scalar glueball, determined solely by the glueball pole mass, and extend the formalism to include excited scalar glueball states through a systematic multipole expansion. The resulting interaction radii, form factors, and projected four-gluon scattering amplitudes are compared with recent lattice Yang--Mills determinations of gravitational form factors~\cite{Abbott:2025irb} and the scalar projection of the four-gluon vertex~\cite{Colaco:2024gmt, Oliveira:2025tfq}. This establishes a direct and systematically improvable connection between an EFT description of scalar glueballs and first-principles lattice Yang--Mills calculations. Future work will extend this framework to tensor and vector glueballs, employing the corresponding projectors developed in lattice QCD~\cite{Aguilar:2024dlv, Aguilar:2023qqd, Colaco:2024gmt, Oliveira:2025tfq} to isolate their contributions, derive their associated form factors and observables, and confront the resulting predictions with lattice determinations.

The manuscript is organized as follows. In \cref{sec:coherent_state}, we motivate the exponential dressing of the gauge field using a coherent-state description of the nonperturbative vacuum. In \cref{glueb}, we derive the effective glueball Lagrangian from the Yang--Mills gluon condensate.  In \cref{glugluon}, we construct the effective glueball--gluon interaction Lagrangian that forms the basis of the present framework. In \cref{greenpro}, we derive the four-gluon Green's function and construct the effective color-singlet projectors. In \cref{extract}, we project the amputated Green's function onto the scalar $J^{PC}=0^{++}$ channel and determine the corresponding effective scattering amplitudes. In \cref{Rformf}, we introduce the reduced and normalized scalar glueball form factors, while in \cref{Peffr} we derive the effective interaction radius and compare it with recent lattice Yang--Mills results. The determination and analysis of the model parameters are presented in \cref{Nanaly}. Finally, our numerical results and concluding remarks are presented in \cref{RN} and \cref{conc}, respectively.

\section{Effective field theory for glueballs}
{\subsection{Coherent-state derivation of the exponential dressing}
\label{sec:coherent_state}
In the subsequent discussion, we introduce a dressed function for the gluon field through
\begin{equation}\label{atz}
A_\mu^a=e^{g_j\phi/\Lambda}\,\tilde{A}_\mu^a,
\end{equation}
where $\tilde{A}_\mu^a$ is the nonperturbative background gauge field associated with the Yang--Mills vacuum condensate, $\phi_j$ is the scalar glueball field, $\Lambda$ is the effective energy scale of the theory, and $g_j$ is a dimensionless coupling constant. This formulation is motivated by the nonperturbative structure of the Yang--Mills vacuum~\cite{Huber:2018ned}. In the background-field formulation \cite{Abbott:1981ke}, the gauge field is decomposed into a nonperturbative background and perturbative fluctuations,
\begin{equation}
A_\mu^a=\tilde{A}_\mu^a+a_\mu^a,
\end{equation}
 where $a_\mu^a$ denotes the perturbative quantum gluon fluctuation. Lorentz invariance and color confinement motivate the identification of the field as the lowest collective excitation. A gauge-invariant color-singlet scalar carrying the vacuum quantum numbers $J^{PC}=0^{++}$, naturally identified with the scalar glueball~\cite{Shifman:1978bx, Shifman:1978by, DiGiacomo:2000irz, Morningstar:1999rf, Chen:2005mg}.

A description of such a collective excitation is provided by the coherent-state formalism~\cite{Glauber:1963tx, Weinberg:1995mt}, which motivates treating the Yang--Mills vacuum as a coherent condensate of scalar glueball quanta. Accordingly, we represent the dressing of the perturbative gluon field by a local analytic dressing function,
\begin{equation}\label{series}
\Phi(x)=1+c_1x+\frac{c_2}{2!}x^2+\cdots,\quad x=\frac{g_j\phi}{\Lambda},
\end{equation}
where $x$ is the dimensionless expansion parameter, with $c_i$ the Taylor-expansion coefficients. 

The coherent-state description of the vacuum motivates the composition law to incorporate a series of scalar glueball excitations in the redefinition of the gauge field. In particular, coherent states are generated by the displacement operator
\begin{equation}
D(\alpha)=\exp\!\left(\alpha a^\dagger-\alpha^{*}a\right),
\end{equation}
where $a$ and $a^\dagger$ denote the annihilation and creation operators of the corresponding collective glueball mode. The displacement operators satisfy the Weyl composition relation \cite{2013qtm..book.....H},
\begin{equation}
D(\alpha_1)D(\alpha_2)=e^{i\,\mathrm{Im}(\alpha_1\alpha_2^*)}D(\alpha_1+\alpha_2),
\end{equation}
where the overall Baker--Campbell--Hausdorff phase is physically irrelevant for the gauge-invariant vacuum expectation values entering the present EFT. Motivated by the Weyl composition law, we require the local dressing function to satisfy
\begin{equation}
\Phi(\phi_1)\Phi(\phi_2)=\Phi(\phi_1+\phi_2).
\end{equation}
Since $\Phi$ is assumed to be analytic, the above functional equation admits the unique solution
\begin{equation}
\Phi(\phi_j)=e^{g_j\phi_j/\Lambda},
\end{equation}
where the normalization conditions $\Phi(0)=1$ and $\Phi'(0)=1$ fix the overall normalization and the linear Taylor coefficient, $c_1$. This yields the field redefinition of \cref{atz}. Because the dressing factor is a gauge-invariant Lorentz scalar, the field redefinition preserves the gauge covariance of the Yang--Mills theory while separating the long-distance nonperturbative background from the perturbative gluonic fluctuations. Although the exponential dressing is not uniquely derived from the fundamental Yang--Mills Lagrangian, it emerges naturally as the effective realization of the coherent-state structure of the nonperturbative Yang--Mills vacuum. 

\subsection{Effective glueball Lagrangian from the gluon condensate}\label{glueb}
Following the discussions in \cref{sec:coherent_state}, we develop the effective glueball--gluon Lagrangian and identify the interactions induced by the nonperturbative Yang--Mills vacuum. The non-Abelian field-strength tensor is defined as
\begin{equation*}
G_{\mu\nu}^a=\partial_\mu A_\nu^a-\partial_\nu A_\mu^a+g\,f^{abc}A_\mu^bA_\nu^c.
\end{equation*}
Substituting \cref{atz}  into the non-Abelian field-strength tensor yields
\begin{align}\label{fstren}
G_{\mu\nu}^a&=e^{g_j \phi/\Lambda}\,{F}_{\mu\nu}^a+\frac{g_j}{\Lambda}e^{g_j \phi/\Lambda}\left(\tilde{A}_\nu^a\,\partial_\mu\phi-\tilde{A}_\mu^a\,\partial_\nu\phi\right)\nonumber\\
&+g\,f^{abc}\tilde{A}_\mu^b\tilde{A}_\nu^c \left(e^{2g_j\phi/\Lambda}-e^{g_j\phi/\Lambda}\right),
\end{align}
where the field-strength tensor associated with the nonperturbative background gauge is
\begin{equation*}
{F}_{\mu\nu}^a=\partial_\mu\tilde{A}_\nu^a-\partial_\nu\tilde{A}_\mu^a+g\,f^{abc}\tilde{A}_\mu^b\tilde{A}_\nu^c.
\end{equation*}
\begin{widetext}
Constructing the squared field-strength tensor from \cref{fstren} yields
\begin{align}
G^a_{\mu\nu} G^{\mu\nu}_a &= e^{2 g_j \phi / \Lambda} \,
{F}^a_{\mu\nu}{F}^{a\,\mu\nu} \nonumber\\
&+ \frac{2 g_j}{\Lambda} e^{2 g_j \phi / \Lambda}{F}^{a}_{\mu\nu}\left(\tilde{A}^{a\,\mu} \partial^\nu \phi - \tilde{A}^{a\,\nu} \partial^\mu \phi\right) \quad + \frac{2 g_j^2}{\Lambda^2} e^{2 g_j \phi / \Lambda}
\left[\left( \tilde{A}^a_\mu \tilde{A}^{a\,\mu} \right)\left( \partial_\nu \phi \, \partial^\nu \phi \right)-\left( \tilde{A}^a_\mu \partial^\mu \phi \right)\left( \tilde{A}^a_\nu \partial^\nu \phi \right)
\right] \nonumber\\
&+ 2 g \, e^{g_j \phi / \Lambda} \left( e^{2 g_j \phi / \Lambda} - e^{g_j \phi / \Lambda} \right)f^{abc} \,{F}^a_{\mu\nu} \tilde{A}^{b\,\mu} \tilde{A}^{c\,\nu}+ g^2 \, f^{abc} f^{ade} \,\tilde{A}^b_\mu \tilde{A}^c_\nu\tilde{A}^{d\,\mu} \tilde{A}^{e\,\nu}\left( e^{2 g_j \phi / \Lambda} - e^{g_j \phi / \Lambda} \right)^2 \nonumber\\
&+ \frac{2 g g_j}{\Lambda} e^{g_j \phi / \Lambda}\left( e^{2 g_j \phi / \Lambda} - e^{g_j \phi / \Lambda} \right)f^{abc}\left(\tilde{A}^a_\nu \partial_\mu \phi- \tilde{A}^a_\mu \partial_\nu \phi\right)\tilde{A}^{b\,\mu} \tilde{A}^{c\,\nu}.
\label{eps1}
\end{align}
We truncate the expansion at second order in the scalar field, retaining all terms up to $\phi^2$. Higher-order terms, $\mathcal{O}(\phi^3)$ and beyond, describe nonlinear glueball self-interactions and higher-dimensional operators, and are neglected within the present EFT framework,
\begin{align}
G^a_{\mu\nu}G^{\mu\nu}_a
&= {F}^a_{\mu\nu}{F}^{a\,\mu\nu}+ \frac{2g_j\phi}{\Lambda}{F}^a_{\mu\nu}{F}^{a\,\mu\nu} + \frac{2g_j^2\phi^2}{\Lambda^2}{F}^a_{\mu\nu}{F}^{a\,\mu\nu}+ \frac{2g_j}{\Lambda}\,
{F}^a_{\mu\nu}\left(\tilde{A}^{a\mu}\partial^\nu\phi - \tilde{A}^{a\nu}\partial^\mu\phi\right)
\nonumber\\
&\quad + \frac{4g_j^2\phi}{\Lambda^2}{F}^a_{\mu\nu}\left(\tilde{A}^{a\mu}\partial^\nu\phi - \tilde{A}^{a\nu}\partial^\mu\phi\right)\frac{2gg_j\phi}{\Lambda}f^{abc}{F}^a_{\mu\nu}\tilde{A}^{b\mu}\tilde{A}^{c\nu} + \frac{5gg_j^2\phi^2}{\Lambda^2}f^{abc}{F}^a_{\mu\nu}\tilde{A}^{b\mu}\tilde{A}^{c\nu}
\nonumber\\
&\quad + \frac{2g_j^2}{\Lambda^2}\left[(\tilde{A}^a_\mu \tilde{A}^{a\mu})(\partial_\nu\phi\,\partial^\nu\phi)- (\tilde{A}^a_\mu \partial^\mu\phi)(\tilde{A}^a_\nu \partial^\nu\phi)\right]
+ \frac{2gg_j}{\Lambda} f^{abc}\left(\tilde{A}^a_\nu \partial_\mu\phi - \tilde{A}^a_\mu \partial_\nu\phi\right)\tilde{A}^{b\mu}\tilde{A}^{c\nu}
\nonumber\\
&\quad + \frac{2gg_j^2\phi}{\Lambda^2} f^{abc}\left(\tilde{A}^a_\nu \partial_\mu\phi - \tilde{A}^a_\mu \partial_\nu\phi\right)\tilde{A}^{b\mu}\tilde{A}^{c\nu} + \frac{g^2 g_j^2 \phi^2}{\Lambda^2}f^{abc} f^{ade}
\tilde{A}^b_\mu \tilde{A}^c_\nu\tilde{A}^{d\mu} \tilde{A}^{e\nu} + \mathcal{O}(\phi^3,\, 1/\Lambda^3).
\end{align}
After integrating by parts and discarding total derivative (surface) terms, which do not contribute to the action under the boundary conditions, the remaining physical interaction terms are,
\begin{align}
G^a_{\mu\nu}G^{\mu\nu}_a&= {F}^a_{\mu\nu}{F}^{a\mu\nu}   \left(1 + \frac{2g_j\phi}{\Lambda}+ \frac{2g_j^2\phi^2}{\Lambda^2}\right)+ \frac{2g_j}{\Lambda}\left[\tilde{A}^a_\mu(\partial_\nu\tilde{A}^{a\nu})
         - \tilde{A}^a_\nu(\partial_\mu\tilde{A}^{a\mu})\right]\partial^\mu\phi \left(1 + \frac{2g_j\phi}{\Lambda}\right)
\nonumber\\
&\quad
  + \frac{2g_j}{\Lambda}\left[\tfrac{1}{2}\tilde{A}^{a\mu}\tilde{A}^a_\mu\Box\phi - \tilde{A}^a_\nu\tilde{A}^{a\mu}\partial_\mu\partial^\nu\phi\right]\left(1 + \frac{2g_j\phi}{\Lambda}\right)- \frac{gg_j\phi}{\Lambda}f^{abc}\tilde{A}^a_\nu(\partial_\mu\tilde{A}^{b\mu})\tilde{A}^{c\nu}\left(1 + \frac{5g_j\phi}{2\Lambda}\right)
\nonumber\\
&\quad
  + \frac{2g_j}{\Lambda}gf^{abc}\tilde{A}^b_\mu\tilde{A}^c_\nu\left(\tilde{A}^{a\mu}\partial^\nu\phi- \tilde{A}^{a\nu}\partial^\mu\phi\right)\left(1 + \frac{2g_j\phi}{\Lambda}\right)+ \frac{2g_j^2}{\Lambda^2}
    \left[(\tilde{A}^a_\mu\tilde{A}^{a\mu})(\partial_\nu\phi\,\partial^\nu\phi)- (\tilde{A}^a_\mu\partial^\mu\phi)(\tilde{A}^a_\nu\partial^\nu\phi)\right]
\nonumber\\
&\quad
    + \frac{7g^2g_j^2\phi^2}{2\Lambda^2}f^{abc}f^{ade}\tilde{A}^b_\mu\tilde{A}^c_\nu\tilde{A}^{d\mu}\tilde{A}^{e\nu} + \mathcal{O}(\phi^3,\,\Lambda^{-3}).
\label{exp2}
\end{align}
\end{widetext}

Taking the vacuum expectation value of \cref{exp2}, and retaining only the leading condensate contributions,
\begin{align}
    \langle G^a_{\mu\nu}G^{\mu\nu}_a\rangle &= \langle{F}^a_{\mu\nu}{F}^{a\mu\nu} \rangle \left[1 + \dfrac{2g_j\phi}{\Lambda}\right]\nonumber\\&+ \dfrac{11g_j^2\phi^2}{2\Lambda^2}\langle{F}^a_{\mu\nu}{F}^{a\mu\nu} \rangle- \dfrac{48g_j^2D^{(2)}}{\Lambda^2}\phi\Box\phi.
\end{align}
To obtain this expression, we have expressed the four-gluon correlator in terms of the gluon condensate through  $\langle {F}^a_{\mu\nu}{F}^{a\mu\nu}\rangle = g^2f^{abc}f^{ade}\langle\tilde{A}^b_\mu\tilde{A}^c_\nu\tilde{A}^{d\mu}\tilde{A}^{e\nu}\rangle$  because the derivatives vanishes. Also, we employ the two-point gluon correlator $\langle \tilde{A}_\mu^a\tilde{A}^{a\mu}\rangle = D^{(2)}\delta^\mu_\mu\delta^{aa} = (4\times 8)\,D^{(2)}$ \cite{Dudal:2005na, Dudal:2008sp, Gubarev:2000eu}, where $D^{(2)}$ denotes the coefficient parameterizing the dimension-two gluon correlator, and $\delta^{aa}=N_A$. For an $\rm SU(N_c)$ symmetry group, $N_A = N_c^2-1$; therefore, for an SU(3) group, $N_A = 8$. To recover the canonical normalization of the scalar kinetic term, we impose
\begin{equation*}
\frac{48g_j^2D^{(2)}}{\Lambda^2}=-\frac{1}{2}\qquad\Longrightarrow\qquad\frac{g_j^2}{\Lambda^2}=-\frac{1}{96D^{(2)}}.
\end{equation*}
Since $g_j^2/\Lambda^2>0$, consistency requires $D^{(2)}<0$. This is not an \textit{ad hoc} assumption: the dimension-two gluon condensate is dynamically generated in Landau-gauge Yang--Mills theory within the Refined Gribov--Zwanziger framework, with its dynamical origin independently established through the Local Composite Operator formalism and its role in the infrared gluon propagator supported by lattice studies~\cite{Dudal:2005na, Dudal:2008sp, Gubarev:2000eu}. In the present Minkowski convention, the corresponding Lorentz-contracted condensate $D^{(2)}=\langle\tilde{A}_\mu^a\tilde{A}^{a\mu}\rangle/32<0$, consistent with the continuation of the positive Euclidean condensate. The sign required for a positive-definite scalar kinetic term therefore coincides with an established nonperturbative feature of Yang--Mills theory. With this choice, the condensate \cite{Shifman:1978by, Gubler:2018ctz} takes the form
\begin{align}\label{eqna}
    \langle G^a_{\mu\nu}G^{\mu\nu}_a\rangle &=\langle F^a_{\mu\nu}F^{a\mu\nu}\rangle\left[1 + \dfrac{2g_j\phi}{\Lambda}\right]\nonumber\\&-\frac{11}{2\times96D^{(2)}}\langle F^a_{\mu\nu}F^{a\mu\nu}\rangle\phi^2+\frac{1}{2}\phi\Box\phi.
\end{align}
Introducing the canonically normalized scalar field, $\phi_R =\phi/2$, and integrating the kinetic term by parts up to a total derivative, the effective Lagrangian assumes the standard form:
\begin{align}\label{effl}
\mathcal L_{\rm eff}=&-\frac14\langle F^a_{\mu\nu}F^{a\mu\nu}\rangle\left[1 + \dfrac{4g_j\phi_R}{\Lambda}\right]+\frac12(\partial_\mu\phi_R)(\partial^\mu\phi_R)\nonumber\\
&-\frac12m_\phi^2\phi_R^2,
\end{align}
where 
\begin{equation}\label{gbmass}
    m_\phi^2=-\frac{11}{96D^{(2)}}\langle F^a_{\mu\nu}F^{a\mu\nu}\rangle=\frac{11g^2_j}{\Lambda^2}\langle F^a_{\mu\nu}F^{a\mu\nu}\rangle.
\end{equation}
We absorb the coupling $g_j$ into the effective scale by redefining
\begin{equation}\label{coplcont}
\frac{g_j}{\Lambda}\equiv\frac{1}{\Lambda_j},
\end{equation}
where $\Lambda_j$ denotes the effective scale associated with the $j$th glueball state. Hereafter, all expressions are written in terms of $\Lambda_j$. The scalar glueball mass emerges dynamically from the nonperturbative Yang--Mills vacuum through the dimension-two and four gluon condensates, rather than being introduced as an independent parameter~\cite{Giacosa:2021brl, Horak:2022aqx}.

\section{Effective field theory of scalar glueball–gluon interactions}\label{glugluon}
After establishing the dynamical generation of the glueball mass from the gluon condensates in \cref{glueb}, we proceed to construct an EFT describing the interactions between scalar glueballs and gluons. The corresponding effective Lagrangian density is given by
\begin{align}\label{lg}
\mathcal{L}=-\sum_j\frac{e^{\phi_j/\Lambda_j}}{4}F_{\mu\nu}^aF_a^{\mu\nu}+\frac{1}{2}\partial_\mu\phi_j\,\partial^\mu\phi^j-\frac{1}{2}\sum_jm_{\phi_j}^2\phi^2_j,
\end{align}
where $F^a_{\mu\nu} = \partial_\mu A^a_\nu-\partial_\nu A_\mu^a + gf^{abc}A_\mu^b A_\nu^c$, is the non-Abelian field-strength tensor, $g$ is the strong coupling and $f^{abc}$ is the structural constant of the SU(3) color group. Expanding the exponential to first order in the scalar field yields
\begin{align}\label{lg1}
\mathcal{L}\supset&-\frac14F_{\mu\nu}^aF_a^{\mu\nu}\left(1+\sum_j\frac{\phi_j}{\Lambda_j}\right)+\frac{1}{2}\partial_\mu\phi_j\,\partial^\mu\phi^j\nonumber\\
&-\frac{1}{2}\sum_jm_{\phi_j}^2\phi^2_j.
\end{align}
which has the same structure as the effective Lagrangian derived in \cref{effl}. 
At leading order, the effective theory describes the interaction of two gluons with a single scalar glueball through the interaction term
\begin{equation}\label{inta}
   \mathcal{L}_{\rm int} \supset-\sum_j\dfrac{\phi_j}{4\Lambda_j} F_{\mu\nu}^a F^{\mu\nu}_a.
\end{equation}
This interaction is used to describe the gluon-fusion process
\begin{equation}\label{proc}
   g(k_1)+g(k_2)\rightarrow\phi\rightarrow g(k_3)+g(k_4), 
\end{equation}
representing gluon-gluon scattering through an intermediate scalar glueball. To derive the interaction vertex in momentum space, we introduce the Fourier decompositions
 \begin{align} \label{frs}
     \phi(x) &= \int \dfrac{d^4 q}{(2\pi)^4}e^{-iqx}\tilde{\phi}(q) \nonumber\\
     A^a_\mu(x) &= \int \dfrac{d^4 k}{(2\pi)^4}e^{-ikx}\tilde{A}^a_\mu(k).
      \end{align}
for the scalar and gluon fields. The Feynman diagram associated with the process \cref{proc} is therefore obtained directly from the leading-order interaction term in \cref{inta}. The $s$, $t$, and $u$ channel diagrams are presented in \cref{fm_diagrams}. 
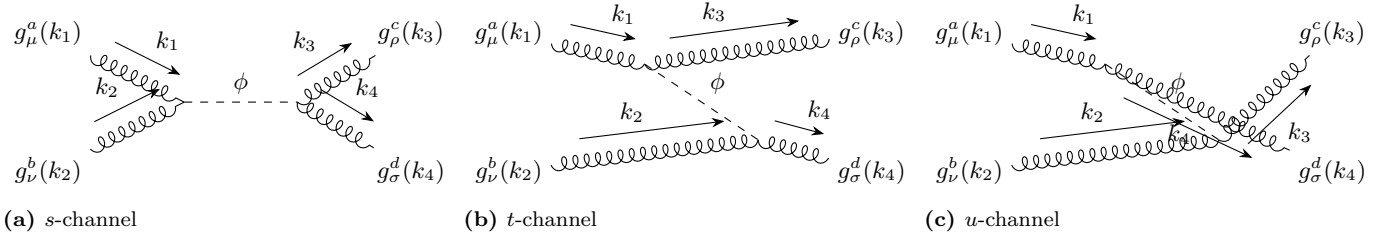
\begin{figure*}[t]
\centering

\begin{subfigure}{0.32\linewidth}
\centering
\begin{tikzpicture}
\begin{feynman}

\vertex (g1) at (-1.8,0.9) {$g^a_\mu(k_1)$};
\vertex (g2) at (-1.8,-0.9) {$g^b_\nu(k_2)$};

\vertex (v1) at (0,0);
\vertex (v2) at (1.5,0);

\vertex (g3) at (3,0.9) {$g^c_\rho(k_3)$};
\vertex (g4) at (3,-0.9) {$g^d_\sigma(k_4)$};

\diagram*{
(g1) -- [gluon, momentum=\(k_1\)] (v1),
(g2) -- [gluon, momentum=\(k_2\)] (v1),
(v1) -- [scalar, edge label=\(\phi\)] (v2),
(v2) -- [gluon, momentum=\(k_3\)] (g3),
(v2) -- [gluon, momentum=\(k_4\)] (g4),
};

\end{feynman}
\end{tikzpicture}
\caption{$s$-channel}
\end{subfigure}
\hfill
\begin{subfigure}{0.32\linewidth}
\centering
\begin{tikzpicture}
\begin{feynman}

\vertex (g1) at (-1.8,0.9) {$g^a_\mu(k_1)$};
\vertex (g2) at (-1.8,-0.9) {$g^b_\nu(k_2)$};

\vertex (v1) at (0,0.5);
\vertex (v2) at (1.5,-0.5);

\vertex (g3) at (3,0.9) {$g^c_\rho(k_3)$};
\vertex (g4) at (3,-0.9) {$g^d_\sigma(k_4)$};

\diagram*{
(g1) -- [gluon, momentum=\(k_1\)] (v1),
(v1) -- [gluon, momentum=\(k_3\)] (g3),
(g2) -- [gluon, momentum=\(k_2\)] (v2),
(v2) -- [gluon, momentum=\(k_4\)] (g4),
(v1) -- [scalar, edge label=\(\phi\)] (v2),
};

\end{feynman}
\end{tikzpicture}
\caption{$t$-channel}
\end{subfigure}
\hfill
\begin{subfigure}{0.32\linewidth}
\centering
\begin{tikzpicture}
\begin{feynman}

\vertex (g1) at (-1.8,0.9) {$g^a_\mu(k_1)$};
\vertex (g2) at (-1.8,-0.9) {$g^b_\nu(k_2)$};

\vertex (v1) at (0,0.5);
\vertex (v2) at (1.5,-0.5);

\vertex (g3) at (3,0.9) {$g^c_\rho(k_3)$};
\vertex (g4) at (3,-0.9) {$g^d_\sigma(k_4)$};

\diagram*{
(g1) -- [gluon, momentum=\(k_1\)] (v1),
(v1) -- [gluon, momentum'={[label distance=-1.5mm]\(k_4\)}] (g4),
(g2) -- [gluon, momentum=\(k_2\)] (v2),
(v2) -- [gluon, momentum'=\(k_3\)] (g3),
(v1) -- [scalar, edge label=\(\phi\)] (v2),
};

\end{feynman}
\end{tikzpicture}
\caption{$u$-channel}
\end{subfigure}
\caption{Feynman diagrams contributing to $gg\rightarrow \phi \rightarrow gg$ scattering mediated by a scalar field $\phi$.}
\label{fm_diagrams}
\end{figure*}

\subsection{Glueball-mediated four-gluon Green's function and $0^{++}$ projection operators}\label{greenpro}
To compute the glueball-mediated contribution to the amputated four-gluon Green's function, we first derive the Euclidean $\phi gg$ vertex associated with the $s$-channel diagram shown in \cref{fm_diagrams}. The corresponding $t$- and $u$-channel vertices are then obtained by the appropriate crossing permutations of the external gluon momenta. The resulting vertex in Euclidean space is
\begin{equation}\label{vetx1}
\Gamma^{ab}_{\mu\nu}(k_1^E,k_2^E)=\frac{1}{\Lambda_j}\delta^{ab}\Big(k_{1\nu}^Ek_{2\mu}^E-(k_1^E\!\cdot\!k_2^E)\delta_{\mu\nu}\Big).
\end{equation}
For convenience, we introduce the Euclidean tensor
\begin{equation}
    T^E_{\mu\nu}(k_i^E,k_j^E)= k_{i\nu}^Ek_{j\mu}^E-k_{i}^E\cdot k_{j}^E\delta_{\mu\nu}.
\end{equation}
Accordingly, for any Mandelstam channel $X=s,t,u$, the $\phi gg$ vertex is
\begin{equation}
\Gamma^{X,E}_{\alpha\beta}(k_i^E,k_j^E)=
\sum_k\frac{1}{\Lambda_k}\,\delta_X\,T_{\alpha\beta}^{E}(k_i^E,k_j^E),
\end{equation}
where $\delta_X$ denotes the color Kronecker delta appropriate to the channel under consideration. The tree-level glueball-exchange amplitudes are obtained by connecting two $\phi gg$ vertices through the scalar glueball propagator and can be written in the compact form
\begin{equation}
\mathcal{M}_X^{E}=\sum_j\Gamma_{j,\,\rm in}^{X,E}\,\frac{1}{X+m_{\phi_j}^2}\,\Gamma_{j,\,\rm out}^{X,E},\qquad X=s,t,u,
\label{eq:amp_compact}
\end{equation}
where $\Gamma_{\mathrm{in}}^{X,E}$ and $\Gamma_{\mathrm{out}}^{X,E}$ denote the incoming and outgoing $\phi gg$ vertices, respectively. Their momentum, color, and Lorentz structures are determined by the corresponding Mandelstam channel, $X$, through the appropriate crossing permutations of the external gluon legs,
\begin{equation}
D_\phi(X)=\sum_{j}\frac{1}{\Lambda^2_j}\frac{1}{X+m_{\phi_j}^2},
\label{eq:glueball_prop}
\end{equation}
is the effective Euclidean scalar glueball propagator. Within the present EFT, the glueball-mediated contribution to the amputated four-gluon Green's function in the $s$-channel takes the form
\begin{align}
\mathcal{G}^{abcd,s}_{\mu\nu\sigma\rho}&=\delta^{ab}\delta^{cd}\,T_{\mu\nu}^E(k_1,k_2)\,D_{\phi}(s)\,T_{\sigma\rho}^E(k_3,k_4), \nonumber\\
\label{Gs_amp}
\end{align}
where the scalar glueball propagator $D_\phi(s)$ is derived from \cref{eq:glueball_prop} at $X=s$. Applying the corresponding crossing permutations to the color and Lorentz structures yields the amputated four-gluon Green's functions in the $t$ and $u$ channels,
\begin{align}
\mathcal{G}^{abcd,t}_{\mu\sigma\nu\rho}&=\delta^{ac}\delta^{bd}\,T_{\mu\sigma}^E(k_1,k_3)\,D_{\phi}(t)\,T_{\nu\rho}^E(k_2,k_4),
\label{Gt_amp}
\\
\mathcal{G}^{abcd,u}_{\mu\rho\sigma\nu}&=\delta^{ad}\delta^{bc}\,T_{\mu\rho}^E(k_1,k_4)\,D_{\phi}(u)\,T_{\sigma\nu}^E(k_2,k_3).
\label{Gu_amp}
\end{align}

To isolate the scalar glueball contribution from the amputated four-gluon Green's function, we introduce a set of color--Lorentz projectors corresponding to the three independent crossing channels. Since the exchanged glueball is a color-singlet, the projectors are obtained by applying the appropriate crossing permutations to both the color and Lorentz indices. Therefore, we construct,
\begin{align}
\mathcal{P}^{(0^{++}),s\,abcd}_{\mu\nu\sigma\rho}&=\delta^{ab}\delta^{cd}\left(\delta_{\mu\nu}\delta_{\sigma\rho}+\delta_{\mu\sigma}\delta_{\nu\rho}+\delta_{\mu\rho}\delta_{\nu\sigma}\right),
\label{eq:proj_scalar_s}
\\
\mathcal{P}^{(0^{++}),t\,abcd}_{\mu\sigma\nu\rho}&=\delta^{ac}\delta^{bd}\left(\delta_{\mu\sigma}\delta_{\nu\rho}+\delta_{\mu\nu}\delta_{\sigma\rho}+\delta_{\mu\rho}\delta_{\sigma\nu}\right),
\label{eq:proj_scalar_t}
\\
\mathcal{P}^{(0^{++}),u\,abcd}_{\mu\rho\sigma\nu}&=\delta^{ad}\delta^{bc}\left(\delta_{\mu\rho}\delta_{\sigma\nu}+\delta_{\mu\sigma}\delta_{\rho\nu}+\delta_{\mu\nu}\delta_{\rho\sigma}\right),
\label{eq:proj_scalar_u}
\end{align}
where the color tensors $\delta^{ab}\delta^{cd}$, $\delta^{ac}\delta^{bd}$, and $\delta^{ad}\delta^{bc}$ project onto the color-singlet exchange in the $s$, $t$, and $u$-channels, respectively, while the same crossing permutations relate the Lorentz structures. The projectors are normalized such that their contraction with the corresponding channel Green's function isolates the scalar color-singlet contribution. The present projectors may be regarded as channel-resolved counterparts of the fully Bose-symmetric tensor $\widetilde{\Gamma}^{(2)}$ introduced in \cite{Colaco:2024gmt, Oliveira:2025tfq}. The projectors $\mathcal{P}^{(0^{++}),s}$, $\mathcal{P}^{(0^{++}),t}$, and $\mathcal{P}^{(0^{++}),u}$ isolate the scalar color-singlet contribution in each Mandelstam channel separately, making them particularly suitable for projecting the amputated four-gluon Green's function onto the physical $0^{++}$ glueball exchange.

\subsection{Projection onto the scalar glueball channel}
\label{extract}
Having constructed the amputated four-gluon Green's function in \cref{Gs_amp,Gt_amp,Gu_amp}, we now project them onto the $J^{PC}=0^{++}$ channel using the projectors defined in \cref{eq:proj_scalar_s,eq:proj_scalar_t,eq:proj_scalar_u}. The resulting quantities correspond to the four-gluon scattering amplitude and contain both the dynamical information associated with scalar glueball exchange and the Lorentz-color tensor structure. The intrinsic scalar glueball form factor will subsequently be extracted by separating these two contributions.

We first consider the $s$-channel, for which the projected scattering amplitude is defined as
\begin{equation}
\mathcal{A}_s(s)=\mathcal{P}^{(0^{++}),s\,abcd}_{\mu\nu\sigma\rho}\mathcal{G}^{abcd,s}_{\mu\nu\sigma\rho}.
\label{eq:Fs_definition}
\end{equation}
Substituting \cref{eq:proj_scalar_s,Gs_amp} into \cref{eq:Fs_definition}, yields
\begin{align}
\mathcal{A}_s(s)=&\delta^{ab}\delta^{cd}\left(\delta_{\mu\nu}\delta_{\sigma\rho}+\delta_{\mu\sigma}\delta_{\nu\rho}+\delta_{\mu\rho}\delta_{\nu\sigma}\right)\times\nonumber\\
&\delta^{ab}\delta^{cd}T_{12}^{\mu\nu}D_\phi(s)T_{34}^{\sigma\rho}.
\end{align}
The color contractions give, $\delta^{ab}\delta^{ab}=\delta^{cd}\delta^{cd}=N_A$ thus,
\begin{equation}
\mathcal{A}_s(s)=N_A^2D_\phi(s)\left[(\mathrm{Tr}\,T_{12})(\mathrm{Tr}\,T_{34})+T_{12}^{\mu\nu}T^{34}_{\mu\nu}+T_{12}^{\mu\nu}T^{34}_{\nu\mu}\right].
\label{eq:Fs_general}
\end{equation}
The Lorentz tensor
\begin{equation}
T_{\mu\nu}(k_i,k_j)=T_{\nu\mu}(k_j,k_i)
\end{equation}
is symmetric under the simultaneous interchange of the Lorentz indices and external gluon momenta, $(\mu,k_i)\leftrightarrow(\nu,k_j)$, reflecting the Bose symmetry of the $\phi gg$ vertex. However, it is not symmetric under the interchange of the Lorentz indices alone, $T_{\mu\nu}(k_i,k_j)\neq T_{\nu\mu}(k_i,k_j)$. For the $s$-channel, we consider general Euclidean external four-momenta $k_i$ satisfying momentum conservation,
\begin{equation}
k_1+k_2=k_3+k_4.
\end{equation}
We define the channel invariants $s=(k_1+k_2)^2$, $t=(k_1-k_3)^2$, and $u=(k_1-k_4)^2$, along with the external virtualities
\begin{equation}
v_i \equiv k_i^2, \qquad i=1,\ldots,4,
\end{equation}
where all scalar products are Euclidean. Momentum conservation gives $s+t+u = \sum_{i=1}^4 v_i$. The required momentum contractions are given by
\begin{align}
k_1\!\cdot\!k_2 &= \tfrac{1}{2}(s-v_1-v_2), & k_3\!\cdot\!k_4 &= \tfrac{1}{2}(s-v_3-v_4), \nonumber \\
k_1\!\cdot\!k_3 &= \tfrac{1}{2}(v_1+v_3-t), & k_2\!\cdot\!k_4 &= \tfrac{1}{2}(v_2+v_4-t), \\
k_1\!\cdot\!k_4 &= \tfrac{1}{2}(v_1+v_4-u), & k_2\!\cdot\!k_3 &= \tfrac{1}{2}(v_2+v_3-u). \nonumber
\end{align}
Using the tensor contractions
\begin{align}
(\mathrm{Tr}\,T_{12})(\mathrm{Tr}\,T_{34}) &= (d-1)^2 (k_1\!\cdot\!k_2)(k_3\!\cdot\!k_4), \nonumber \\
T_{12}^{\mu\nu}T^{34}_{\mu\nu} &= (d-2)(k_1\!\cdot\!k_2)(k_3\!\cdot\!k_4) \nonumber \\
&\quad + (k_1\!\cdot\!k_3)(k_2\!\cdot\!k_4), \\
T_{12}^{\mu\nu}T^{34}_{\nu\mu} &= (d-2)(k_1\!\cdot\!k_2)(k_3\!\cdot\!k_4) \nonumber \\
&\quad + (k_1\!\cdot\!k_4)(k_2\!\cdot\!k_3), \nonumber
\end{align}
where $d=4$ is the spacetime dimension. The projected $s$-channel amplitude becomes
\begin{align}
\mathcal{A}_s(s,t,u) &= \frac{N_A^2}{4} D_\phi(s) \Big[ (d^2-3)(s-v_1-v_2)(s-v_3-v_4) \nonumber \\
&\quad + (v_1+v_3-t)(v_2+v_4-t) \nonumber \\
&\quad + (v_1+v_4-u)(v_2+v_3-u) \Big].
\label{eq:Fs_final}
\end{align}
For comparison with the conventional physical $gg\rightarrow\phi\rightarrow gg$ scattering amplitude, one may subsequently consider the formal massless on-shell limit $v_i \rightarrow 0$. In this limit, $u+t+s\rightarrow0$ and \cref{eq:Fs_final} reduces to 
\begin{equation}
\mathcal{A}_s(s)=N_A^2\left[\frac{d^2-3}{4}s^2+\frac14(t^2+u^2)\right]D_\phi(s),
\label{eq:Fs_final}
\end{equation}
which represents the scalar projection of the amputated four-gluon Green's function in the $s$-channel. The projected amplitudes in the crossed channels are obtained by applying the corresponding crossing transformations,
\begin{align}
(s,12,34)\rightarrow (t,13,24),\quad (s,12,34)\rightarrow(u,14,23),
\end{align}
together with the Green's functions \cref{Gt_amp,Gu_amp}. Since the color contractions are identical to those of the $s$-channel, the projected amplitudes become
\begin{align}
\mathcal{A}_t(t)&=N_A^2\left[\frac{d^2-3}{4}t^2+\frac14(s^2+u^2)\right]D_\phi(t)\label{f1},
\\
\mathcal{A}_u(u)&=N_A^2\left[\frac{d^2-3}{4}u^2+\frac14(s^2+t^2)\right]D_\phi(u)\label{f2}.
\end{align}
The amplitudes $\mathcal{A}_s$, $\mathcal{A}_t$, and $\mathcal{A}_u$ contain both the universal kinematic factors arising from the Lorentz and color projections and the dynamical information associated with scalar glueball exchange through the propagator $D_\phi(X)$. 

\subsection{Reduced and normalized scalar glueball form factor}\label{Rformf}
The projected quantities obtained in \cref{extract}, $\mathcal{A}_s(s)$, $\mathcal{A}_t(t)$, and $\mathcal{A}_u(u)$, are the physical scattering amplitudes predicted by the EFT. Inspection of \cref{eq:Fs_final,f1,f2} shows that each projected amplitude factorizes naturally into a scalar function and a channel-dependent kinematic coefficient,
\begin{equation}
\mathcal{A}_X(X)=K_X(s,t,u)\,\mathcal G_\phi(X),
\label{eq:FXfactorization}
\end{equation}
where
\begin{align}
K_s &=N_A^2\left[\frac{d^2-3}{4}s^2+\frac14(t^2+u^2)\right],\\
K_t &=N_A^2\left[\frac{d^2-3}{4}t^2+\frac14(s^2+u^2)\right],\\
K_u &=N_A^2\left[\frac{d^2-3}{4}u^2+\frac14(s^2+t^2)\right],
\end{align}
depend exclusively on the Lorentz contractions, color algebra, and the choice of projector, whereas
\begin{equation}
\mathcal G_\phi(X)=D_\phi(X)=\sum_{j=0}^{n}\frac{1}{\Lambda^2_j}\frac{1}{X+m_{\phi_j}^2},
\label{eq:Gphi}
\end{equation}
contains the scalar glueball dynamics. Within the present tree-level effective theory, $\mathcal G_\phi(X)$ coincides with the effective scalar glueball propagator introduced in \cref{eq:glueball_prop}. 

To isolate the intrinsic momentum dependence of the exchanged scalar glueball, we define the reduced glueball form factor
\begin{equation}
\mathcal F_\phi(X) \equiv\frac{\mathcal{A}_X(X)}{K_X(s,t,u)}=\mathcal G_\phi(X).
\label{eq:ReducedFF}
\end{equation}
This definition removes the purely kinematic contributions arising from the projection procedure while preserving the complete dynamical content of the scalar glueball exchange. The reduced form factor is normalized at vanishing momentum transfer, following the convention commonly adopted in lattice determinations of glueball gravitational form factors~\cite{Abbott:2025irb, Chen:2005mg}. The reduced form factor is normalized at vanishing momentum transfer,
\begin{equation}
\widehat{\mathcal F}_\phi(X)=\frac{\mathcal F_\phi(X)}{\mathcal F_\phi(0)},\qquad \widehat{\mathcal F}_\phi(0)=1.
\label{eq:NormalizedFF}
\end{equation}
The normalization removes the overall interaction strength without altering the momentum dependence. Although the overall coupling is eliminated, the normalized form factor continues to depend on the relative effective couplings of the exchanged glueball states in the multipole expansion, thereby enabling meaningful comparisons among different theoretical approaches.
Since the effective interaction is crossing symmetric, it is convenient to introduce a crossing-symmetric normalized one-glueball form factor,
\begin{equation}
\widehat{\mathcal F}_{\mathrm{sym}}(s,t,u) =\frac13\left[\widehat{\mathcal F}_\phi(s)+\widehat{\mathcal F}_\phi(t)+\widehat{\mathcal F}_\phi(u)\right],
\label{eq:Fsym}
\end{equation}
which satisfies
\begin{equation}
\widehat{\mathcal F}_{\mathrm{sym}}(0,0,0)=1.
\end{equation}
Therefore, for a multipole glueball state,
\begin{equation}
\widehat{\mathcal F}_{\rm sym}(s,t,u)=\frac{\displaystyle\sum_{j=0}^{n}\dfrac{1}{\Lambda_j^2}\left(\frac{1}{s+m_{\phi_j}^2}+\frac{1}{t+m_{\phi_j}^2}+\frac{1}{u+m_{\phi_j}^2}\right)}{\displaystyle3\sum_{j=0}^{n}\frac{1}{m_{\phi_j}^2\Lambda^2_j}}.
\label{eq:Fsym_explicit}
\end{equation}
The quantity $\widehat{\mathcal F}_{\mathrm{sym}}$ provides a crossing-symmetric measure of the intrinsic scalar glueball structure, independent of the projector-dependent Lorentz and color factors. 

The crossing-symmetric form factor in \cref{eq:Fsym_explicit} can be viewed, in analogy with resonance-saturation approaches such as vector meson dominance (VMD), as arising from the exchange of a spectrum of scalar-glueball states~\cite{Sakurai:1960ju, Sakurai:1969ss}. In the present case, the scalar-glueball dynamics entering each Mandelstam channel is described by the same spectral function of the invariant momentum carried by the exchanged state. This motivates introducing the corresponding one-variable normalized form factor,
\begin{equation}\label{tawff}
\widehat{\mathcal F}_{\phi}(q^2)=\sum_j c_j\,\frac{m_{\phi_j}^2}{q^2+m_{\phi_j}^2},
\end{equation}
where $q^2$ denotes the squared momentum flowing through the exchanged scalar-glueball propagator. For the three crossing channels, we evaluate the argument independently as $q^2=s$, $t$, or $u$; in general, these invariants differ for a given kinematic configuration, as derived in \cref{eq:Fsym}. Therefore, the one-variable representation in \cref{tawff} does not assume equality of the three Mandelstam variables, but instead provides the common dynamical function evaluated separately in each channel.

The effective interaction scales and glueball masses fix the dimensionless coefficients $c_j$,
\begin{equation}
c_j \equiv
\frac{\displaystyle {\Lambda_j^{-2} m_{\phi_j}^{-2}}}
{\displaystyle \sum_{k=0}^{n}{\Lambda_k^{-2} m_{\phi_k}^{-2}}},
\label{eq:cj}
\end{equation}
where the index $j$ labels the individual scalar glueball state, while the summation index $k$ runs over all states included in the multipole expansion. By construction,
\begin{equation}
\sum_{j=0}^{n} c_j = 1,
\end{equation}
which ensures the normalization conditions. Thus, the masses $m_{\phi_j}$ determine the momentum dependence associated with the individual glueball poles, while the coefficients $c_j$ encode their relative contributions to the normalized scalar-glueball response. {The form factor \cref{tawff} is associated with a dressing of a scalar density operator, in analogy with the vector meson dominance in the dressed vector current. This association is important for comparing our results with lattice calculations of the gravitational form factors of the ground-state glueball~\cite{Abbott:2025irb}, as we present in the following sections.}

\section{Prediction of the effective interaction radius}\label{Peffr}
To assess the predictive power of the proposed EFT, we estimate the scalar glueball interaction radius and compare it with the Breit-frame mass-radius extracted from the lattice Yang--Mills gravitational form factors in \cite{Abbott:2025irb}. Given the possibility of mixing between glueball and conventional hadronic states, the unique identification of pure glueball states remains challenging. Determining their characteristic size can therefore provide a complementary observable for distinguishing glueballs from conventional hadrons. Several hadronic models \cite{Schafer:1994fd, Forkel:2000fd, Hou:2002jv}, for example, predict that glueballs can be more compact than conventional hadrons, whose characteristic size is set by the nonperturbative QCD scale, $r_{\rm had}\sim 1/\Lambda_{\rm QCD}$. 

Similar to the conventional definition of the electromagnetic charge radius extracted from the slope of a normalized form factor at vanishing momentum transfer \cite{Gao:2021sml, Sick:2018fzn, PhysRev.126.2256, Polyakov:2018zvc, Freese:2019bhb}, we define the effective interaction radius as  
\begin{equation}
r_{\rm int}^2=-6\left.\frac{1}{F(q^2)}\frac{dF(q^2)}{dq^2}\right|_{q^2=0}.
\end{equation}
For the scalar-glueball contribution, the radius is defined from the one-variable normalized form factor $\widehat{\mathcal F}_{\phi}(q^2)$ introduced in \cref{tawff}, rather than directly from the three-variable crossing-symmetric quantity $\widehat{\mathcal F}_{\rm sym}(s,t,u)$. This distinction is important because $s$, $t$, and $u$ are generally different for a given kinematic configuration. The one-variable form factor is evaluated independently in each channel, with $q^2=s$, $t$, or $u$, while the crossing-symmetric form factor is obtained from their symmetric combination. 

Since $\widehat{\mathcal F}_{\phi}(0)=1$, we therefore define
\begin{equation}\label{dform}
r_{\rm int}^2=-6\left.\frac{d\widehat{\mathcal F}_{\phi}(q^2)}
{dq^2}\right|_{q^2=0}.
\end{equation}
For a single glueball pole, the one-variable form factor reduces to
\begin{equation}
\widehat{\mathcal F}_{\phi}(q^2)=\frac{m_\phi^2}{q^2+m_\phi^2},
\end{equation}
which yields
\begin{equation}
\left.\frac{d\widehat{\mathcal F}_{\phi}(q^2)}{dq^2}\right|_{q^2=0}=-\frac{1}{m_\phi^2}.
\end{equation}
The effective interaction radius therefore follows as
\begin{equation}
r_{\rm int}^2=\frac{6}{m_\phi^2} \qquad\rightarrow\qquad r_{\rm int}=\frac{\sqrt{6}}{m_\phi} .
\label{eq:rint}
\end{equation}
 The corresponding predictions are summarized in \cref{tab:scalar_glueball}, illustrating the inverse correlation $r_{\rm int}\propto 1/m_\phi$. For a given glueball mass, the interaction radius is determined solely by the pole structure of the normalized one-variable form factor and is therefore independent of the overall effective interaction scale $\Lambda$ and the gluon condensate. The glueball candidates $f_0(1370)$, $f_0(1500)$, and $f_0(1710)$ predict $r_{\rm int}$ of $0.36$, $0.32$, and $0.28$ fm, respectively. The prediction for $f_0(1710)$ is numerically consistent with the recently reported lattice Yang--Mills mass-radius, $0.263(31)$ fm~\cite{Abbott:2025irb}. Although the effective interaction radius and the Breit-frame mass-radius extracted from gravitational form factors are distinct observables, their comparable numerical values suggest that the pole structure of the present EFT captures a characteristic spatial scale of the scalar glueball.

 The comparison with the lattice gravitational form factor is motivated by their common dependence on the invariant momentum transfer. The lattice calculation determines the gravitational form factors $A(t)$ and $D(t)$ from the energy-momentum tensor, with $A(t)$ governing the energy-momentum distribution and $D(t)$ encoding additional mechanical information such as internal stresses. Since the lattice mass radius is determined primarily by the slope of $A(t)$, with $D(0)$ entering as a finite-mass correction, the normalized $A(t)/A(0)$ provides the most natural lattice counterpart to our normalized scalar-glueball form factor. Furthermore, the trace of the energy-momentum-tensor matrix element contains a leading contribution proportional to the glueball mass squared through $A(t)$, whereas the $D(t)$ contribution enters with an explicit factor of the momentum transfer ~\cite{Abbott:2025irb}. Given the glueball mass scale, this further supports $A(t)$ as the appropriate lattice quantity for the present comparison. Therefore, we compare the momentum-transfer dependence of our form factor to $A(t)$ and use the corresponding lattice mass radius as an independent benchmark for our interaction radius.
 
 
 For a multipole spectrum of scalar glueball states, the same definition applied to the one-variable form factor $\widehat{\mathcal F}_{\phi}(q^2)$ gives
\begin{equation}
r_{\rm int}^2=6\frac{\displaystyle\sum_{j=0}^{n}{m_{\phi_j}^{-4}\Lambda_j^{-2}}}{\displaystyle\sum_{j=0}^{n}{m_{\phi_j}^{-2}\Lambda_j^{-2}}}.
\label{eq:rint_excited}
\end{equation}
For two states, $j=0,1$, this becomes
\begin{equation}
r_{\rm int}^{*2}=\frac{6}{m_{\phi_0}^2m_{\phi_1}^2}\,\frac{m_{\phi_1}^4\Lambda_1^2+m_{\phi_0}^4\Lambda_0^2}{m_{\phi_1}^2\Lambda_1^2+m_{\phi_0}^2\Lambda_0^2}.
\label{eq:rint_two}
\end{equation}

\subsection{Parameter determination and numerical results}\label{Nanaly}
\begin{table}[t!]
\centering
\begin{tabular}{lcccc}
\toprule
State & Mass  & $r_{\rm int}$ & Glueball & Included \\
& (GeV) & (fm)&likelihood&  in fit \\
\midrule
$f_0(500)$  & 0.55 & 0.88 & Very unlikely & No \\
$f_0(980)$  & 0.99 & 0.49 & Very unlikely & No \\
$f_0(1370)$ & 1.37 & 0.36 & Possible      & Yes \\
$f_0(1500)$ & 1.50 & 0.32 & Strong        & Yes \\
$f_0(1710)$ & 1.73 & 0.28 & Strongest     & Yes \\
$f_0(2020)$ & 1.99 & 0.24 & Weak          & No \\
$f_0(2100)$ & 2.10 & 0.23 & Weak          & No \\
$f_0(2200)$ & 2.19 & 0.22 & Weak          & No \\
$f_0(2330)$ & 2.33 & 0.21 & Weak          & No \\
\bottomrule
\end{tabular}
\caption{Scalar-isoscalar mesons listed by the PDG \cite{PhysRevD.110.030001}. The third column gives the effective interaction radius predicted by the present EFT, obtained from the slope of the normalized scalar glueball form factor according to \cref{eq:rint}. The fourth column summarizes the current phenomenological and lattice-QCD assessment of each state as a scalar glueball candidate (see, e.g., Refs.~\cite{Amsler:1995tu, Close:2000yk, Cheng:2006hu, Mathieu:2008me, Ren:2023ebq}), while the final column indicates the states included in the present analysis.}
\label{tab:scalar_glueball}
\end{table}

\begin{table*}[t!]
\centering
\begin{tabular}{llccccc}
\toprule
&&\multicolumn{5}{c}{Predicted $\Lambda_j$ (GeV)}\\
\cmidrule(lr){3-7}
Method &
$\left\langle\frac{\alpha_s}{\pi}G^2\right\rangle$ (GeV$^4$)&$f_0(1370)$&$f_0(1500)$&$f_0(1710)$&Pure YM $0^{++}$&First excited $0^{++}$\\

&&1.37 GeV&1.50 GeV&1.73 GeV \cite{Morningstar:1999rf,Chen:2005mg}&1.73 GeV \cite{Cheng:2015iaa}&2.67 GeV \cite{Morningstar:1999rf}\\

&&$\alpha_s=0.362$&$\alpha_s=0.343$&$\alpha_s=0.317$&$\alpha_s=0.317$&$\alpha_s=0.259$\\

\midrule
Phenomenological QCD sum rules &$0.0120\pm0.0040$ \cite{Shifman:1978bx,Shifman:1978by} &0.781&0.733&0.661&0.661&0.474\\

Lattice QCD + NSPT &$0.0280\pm0.0030$ \cite{Horsley:2012ra} &1.193&1.120&1.010&1.010&0.724\\


QCD sum rules &$0.0202\pm0.0011$ \cite{Narison:2023npv} &1.014&0.951&0.858&0.858&0.615\\

QCD sum rules &$0.0239\pm0.0080$ \cite{Narison:2023npv} &1.103&1.035&0.933&0.933&0.669\\

QCD sum rules &$0.0203\pm0.0011$ \cite{Narison:2018dcr,Narison:2018xbj} &1.016&0.953&0.860&0.860&0.617\\

QCD sum rules &$0.0271\pm0.0095$ \cite{Narison:2018dcr,Narison:2018xbj} &1.174&1.102&0.994&0.994&0.712\\

QCD sum rules &$0.0234\pm0.0021$ \cite{Narison:2018dcr,Narison:2018xbj} &1.091&1.024&0.923&0.923&0.662\\

QCD sum rules &$0.0199\pm0.0014$ \cite{Narison:2018dcr} &1.006&0.944&0.851&0.851&0.610\\

\bottomrule
\end{tabular}
\caption{Predicted effective scales $\Lambda_j$, obtained from \cref{eq:Lambda_extract} using representative determinations of the vacuum gluon condensate. The first three columns after the columns of the gluon condensates correspond to the phenomenological scalar glueball candidates, while the last two columns show the lattice Yang--Mills ground-state and first excited $0^{++}$ glueballs with masses $m_{\phi_0}=1.73$ GeV and $m_{\phi_1}=2.67$ GeV, respectively.}
\label{tab:lambda}
\end{table*}

\begin{table}[t]
\centering
\begin{tabular}{llcc}
\toprule
$\left\langle\frac{\alpha_s}{\pi}G^2\right\rangle$
(GeV$^4$) &$\Lambda_0$ (GeV) &$\Lambda_1$ (GeV) &$r_{\rm int}^*$ (fm) \\
\midrule

$0.0120\pm0.0040$  & 0.661 & 0.474 & 0.24022 \\

 $0.0280\pm0.0030$  & 1.010 & 0.724 & 0.24020 \\


 $0.0202\pm0.0011$ & 0.858 & 0.615 & 0.24020 \\

 $0.0239\pm0.0080$ & 0.933 & 0.669 & 0.24021 \\

 $0.0203\pm0.0011$ & 0.860 & 0.617 & 0.24024 \\

 $0.0271\pm0.0095$ & 0.994 & 0.712 & 0.24017 \\

 $0.0234\pm0.0021$ & 0.923 & 0.662 & 0.24023 \\

 $0.0199\pm0.0014$ & 0.851 & 0.610 & 0.24020 \\

\bottomrule
\end{tabular}
\caption{Predicted effective scales for the ground-state ($\Lambda_0$) and first excited ($\Lambda_1$) scalar glueballs, together with the corresponding two-state interaction radius $r_{\rm int}^*$ obtained from \cref{eq:rint_two}. Although the extracted effective scales depend on the gluon-condensate determination, the interaction radius remains remarkably stable, varying by less than $0.03\%$ across all representative inputs. This demonstrates that the predicted reduction of the interaction radius due to the first excited scalar glueball is largely insensitive to the choice of gluon condensate.}
\label{tab:lambda1}
\end{table}

\begin{table}[t!]
\centering
\begin{tabular}{lcccccc}
\toprule
Candidate & $m_\phi$ & $N$ & $\chi^2$ & $\chi_{\rm red}^2$ & $\Delta\chi^2$ & data \\
& (GeV) & & & & &points\\
\midrule
$f_0(1370)$ & 1.37 & $1.111 \pm 0.025$ & 18.865 & 1.451 & 10.977 & 14 \\
$f_0(1500)$ & 1.50 & $1.066 \pm 0.024$ & 12.263 & 0.943 & 4.376 & 14 \\
$f_0(1710)$ & 1.73 & $1.005 \pm 0.023$ & 7.888  & 0.607 & 0.000 & 14 \\
\bottomrule
\end{tabular}
\caption{Best-fit normalization factor $N$, chi-square ($\chi^2$), reduced chi-square ($\chi_{\rm red}^2$), and relative chi-square difference ($\Delta\chi^2=\chi^2-\chi^2_{\rm min}$) obtained by fitting the normalized monopole form factor to the lattice Yang--Mills scalar glueball gravitational form factor. Since all three hypotheses are fitted to the same dataset with the same number of free parameters, $\Delta\chi^2$ provides a direct measure of their relative agreement with the lattice data. The fitted normalization approaches unity and the fit quality improves systematically with increasing glueball mass, with the $f_0(1710)$ candidate yielding the lowest $\chi^2$ and $\Delta\chi^2=0$, thereby providing the strongest agreement with the lattice momentum dependence.}
\label{tab:chi2}
\end{table}

\begin{table}[t]
\centering
\begin{tabular}{lcccc}
\toprule
Configuration & External momenta & $s$ & $t$ & $u$ \\
\midrule
Config.~1 & $(p,\,p,\,p,\,-3p)$ & $4p^2$ & $0$ & $16p^2$ \\
Config.~2 & $(0,\,p,\,2p,\,-3p)$ & $p^2$ & $4p^2$ & $9p^2$ \\
Config.~3 & $(0,\,p,\,p,\,-2p)$ & $p^2$ & $p^2$ & $4p^2$ \\
\bottomrule
\end{tabular}
\caption{Mandelstam invariants corresponding to the three kinematic configurations used in the evaluation of the crossing-symmetric scalar glueball form factor. The external momenta are chosen as collinear Euclidean four-momenta parameterized by the common momentum scale $p$ satisfying momentum conservation, with the Mandelstam variables defined as $s=(p_1+p_2)^2$, $t=(p_1-p_3)^2$, and $u=(p_1-p_4)^2$.}
\label{tab:kinematics}
\end{table}

In this section, we determine the model parameters and use them to compute the glueball form factors numerically. To determine $\Lambda_j$, we first evaluate the strong coupling $\alpha_s$ at the characteristic glueball mass using the two-loop pQCD running coupling. The running coupling at the renormalization scale $Q$ is given by
\begin{equation}
\alpha_s(Q^2)=\frac{4\pi}{\beta_0L}\left[1-\frac{\beta_1}{\beta_0^2}\frac{\ln L}{L}\right],\quad L=\ln\left(\frac{Q^2}{\Lambda_{\rm QCD}^2}\right),
\label{eq:alphaS_twoloop}
\end{equation}
where
\begin{equation}
\beta_0=11-\frac{2}{3}n_f, \qquad \beta_1=102-\frac{38}{3}n_f,
\label{eq:beta_coeff}
\end{equation}
are the one and two-loop coefficients of the QCD $\beta$ function, respectively \cite{Deur:2016tte}. Throughout this work, we identify the renormalization scale with the glueball mass, $Q=m_\phi$, and adopt $\Lambda_{\rm QCD}=0.3~\mathrm{GeV}$. Although the effective theory contains only glueball degrees of freedom, the running of the strong coupling is inherited from the underlying QCD theory. We therefore evaluate $\alpha_s$ using the two-loop QCD $\beta$ function with $n_f=4$, appropriate for the renormalization scales $Q\simeq1.4$--$1.7~\mathrm{GeV}$ considered here. Since phenomenological and lattice determinations of the gluon condensate are quoted at comparable renormalization scales \cite{Deur:2023dzc}, this provides a consistent matching for extracting $\Lambda_j$. The resulting values of $\alpha_s(m_\phi^2)$ are listed in \cref{tab:lambda}.

The $\Lambda_j$ follows from \cref{gbmass} as
\begin{equation}
\Lambda_j=\sqrt{\frac{11}{m_{\phi_j}^2}\left\langle\frac{\alpha_s}{\pi}G_{\mu\nu}^aG^{a\mu\nu}\right\rangle\frac{\pi}{\alpha_s(m_{\phi_j}^2)}},
\label{eq:Lambda_extract}
\end{equation}
where we have used the identity
\begin{equation}\label{condst}
\left\langle F_{\mu\nu}^aF^{a\mu\nu}\right\rangle=\frac{\pi}{\alpha_s(m_{\phi_j}^2)}\left\langle\frac{\alpha_s}{\pi}G_{\mu\nu}^aG^{a\mu\nu}\right\rangle,
\end{equation}
which relates the conventionally quoted gluon condensate to the gluon field-strength condensate. 

Lattice QCD predicts a tower of scalar glueball states with quantum numbers $J^{PC}=0^{++}$, comprising a well-established ground state and a sequence of excited resonances. In the pioneering anisotropic lattice study of \cite{Morningstar:1999rf}, the scalar glueball spectrum yields a ground-state mass of $m_{\phi_0}=1730\pm50\pm80~\mathrm{MeV}$ and a first excited state of $m_{\phi_1}=2670\pm180\pm130~\mathrm{MeV}$. Subsequent lattice calculations confirmed the ground-state mass near $1.7~\mathrm{GeV}$ and further refined the glueball spectrum~\cite{Chen:2005mg, Li:2013pta}. These results provide a natural motivation for extending the present single-pole EFT to include excited scalar glueball states through a multi-pole expansion of the form factor. The coupling scale and the effective interaction radius for the two-state glueballs for the corresponding gluon condensates have been reported in \cref{tab:lambda1}.

Compared to $r_{\rm int}=\sqrt{6}/m_{\phi_0}$, which is parameter-free, the two-state result depends explicitly on $\Lambda_j$, \cref{eq:rint_excited}. Although the first excited state has a larger effective inverse coupling ($1/\Lambda_1^2>1/\Lambda_0^2$) (see \cref{tab:lambda}), its contribution is suppressed by its larger mass ($m_{\phi_1}>m_{\phi_0}$), leaving the interaction radius largely controlled by the ground state. Using $m_{\phi_0}=1.73$ GeV, $m_{\phi_1}=2.67$ GeV, and the corresponding gluon condensates from \cref{tab:lambda}, we obtain $r_{\rm int}^*=0.240\pm 0.00003$ (see \cref{tab:lambda1}) fm across all the condensates, compared to the ground-state value of $0.279$ fm. The resulting $\sim14\%$ reduction indicates that excited scalar glueballs provide only a modest short-range correction while the interaction radius remains dominated by the ground-state pole. The two-state prediction is also compatible with the recent lattice determination, $r=0.263(31)$ fm, differing by only $\sim0.7\sigma$. The slight downward shift relative to the lattice central value is therefore a prediction of the multipole extension of the present EFT and provides a quantitative benchmark for future lattice calculations of the gravitational form factors of excited $0^{++}$ glueball states.

Furthermore, the values obtained for $\Lambda_j$, ranging from approximately $0.66$ to $1.2~\mathrm{GeV}$ (from \cref{tab:lambda}), are consistent with the characteristic nonperturbative energy scale of QCD generated through dimensional transmutation~\cite{Deur:2016tte}. They are significantly larger than the fundamental parameter $\Lambda_{\rm QCD}\sim0.2\text{--}0.3~\mathrm{GeV}$, while remaining within the characteristic nonperturbative QCD scale set by the confinement scale, $\sqrt{\sigma}\simeq0.42~\mathrm{GeV}$, and the lightest scalar-glueball mass, $m_{0^{++}}\sim1.7~\mathrm{GeV}$~\cite{Bali:2000gf, Lucini:2004my, Morningstar:1999rf, Chen:2005mg}. This supports the interpretation of $\Lambda_j$ as an effective interaction scale governing scalar glueball--gluon dynamics within the present EFT.

To determine the dynamically generated $m_\phi$ from \cref{gbmass}, we take the dimension-two gluon condensate $g^2\langle A_\mu^a A^{a\mu}\rangle_E$ as the nonperturbative input and determine the strong coupling self-consistently at the dynamically generated scale $Q^2=m_\phi^2$ using the two-loop running coupling in \cref{eq:alphaS_twoloop,eq:beta_coeff}. The coupling is related to the usual strong coupling through $g^2(m_\phi^2)=4\pi\alpha_s(m_\phi^2)$. Since the dimension-two condensate relates $g^2(m_\phi^2)$ to the effective interaction scale, $\Lambda_j = -96\,D^{(2)}$, and $\alpha_s$ itself depends on $m_\phi$ through the renormalization-group equation. The quantities $m_\phi$, $\alpha_s(m_\phi^2)$, and $g(m_\phi^2)$ form a coupled system that must be solved simultaneously. We adopt the Euclidean convention $g^2\langle A_\mu^a A^{a\mu}\rangle_E>0$, for the dimension-two condensate quoted in the literature \cite{Dudal:2010tf,RuizArriola:2004en,Boucaud:2008gn}. In accordance with the Euclidean-to-Minkowski continuation adopted in \cref{glueb}, the corresponding Minkowski contraction is negative, $\langle A_\mu^a A^{a\mu}\rangle_M=\langle A_0^{a\,2}\rangle-\langle A_j^{a\,2}\rangle<0$, and hence, with the discussion preceding \cref{eqna},
\begin{equation*}
D^{(2)}_M =D^{(2)}=\frac{\langle A_\mu^a A^{a\mu}\rangle_M}{32}<0.
\end{equation*}
We then use representative determinations of the dimension-two condensate together with the fixed dimension-four gluon condensate to obtain the self-consistent parameter sets, $m_\phi$, $\alpha_s(m_\phi^2)$, and $g(m_\phi^2)$, summarized in \cref{tab:self_consistent}.

\begin{table*}[t!]
\centering
\begin{tabular}{lcccccccc}
\toprule
$\left\langle g^2 A^2\right\rangle_E$& $\left\langle \dfrac{\alpha_s}{\pi}G_{\mu\nu}^aG^{a\mu\nu}\right\rangle$& $m_\phi$& $\alpha_s(m_\phi^2)$& $g(m_\phi^2)$& $\langle A^2\rangle_E$& $D_M^{(2)}$& $\Lambda_j$\\

[GeV$^2$] & [GeV$^4$] & [GeV] &  &  & [GeV$^2$] & [GeV$^2$] & [GeV]\\
\midrule

$3.03$ \cite{Dudal:2010tf} & $0.0280$ & $1.157$ & $0.407$ & $2.261$ & $0.593$ & $-0.019$ & $1.333$ \\

$3.29$ \cite{Dudal:2010tf} & $0.0280$ & $1.110$ & $0.420$ & $2.297$ & $0.624$ & $-0.019$ & $1.368$\\

$4.40$ \cite{RuizArriola:2004en} & $0.0280$ & $0.960$ & $0.474$ & $2.442$ & $0.738$ & $-0.023$ & $1.488$ \\

$5.10$ \cite{Boucaud:2008gn} & $0.0280$ & $0.891$ & $0.509$ & $2.530$ & $0.797$ & $-0.025$ & $1.546$ \\
\bottomrule
\end{tabular}
\caption{Self-consistent glueball parameters obtained for representative values of the dimension-two gluon condensate, with the dimension-four gluon condensate fixed at $\left\langle(\alpha_s/\pi)G_{\mu\nu}^aG^{a\mu\nu}\right\rangle=0.0280~\mathrm{GeV}^4$.}
\label{tab:self_consistent}
\end{table*}

\section{Results and analysis}\label{RN}
The results summarized in \cref{tab:scalar_glueball,tab:lambda,tab:lambda1} highlight the complementary roles of the glueball mass and the gluon condensate within the present EFT. As shown in \cref{tab:scalar_glueball}, the $r_{\rm int}$ decreases monotonically with increasing glueball mass, following the parameter-free relation $r_{\rm int}\propto 1/m_\phi$. The favored glueball candidates $f_0(1370)$, $f_0(1500)$, and $f_0(1710)$ are predicted to have effective interaction radii of $0.36$, $0.32$, and $0.28$ fm, respectively. The prediction for $f_0(1710)$ is in excellent agreement with the lattice Yang--Mills mass-radius, $0.263(31)$ fm~\cite{Abbott:2025irb}. Moreover, the $r_{\rm int}$ for the tabulated glueball masses with $m_\phi\geq 1.73$ GeV, are broadly consistent with the estimates from lattice QCD and Bethe--Salpeter approaches, which typically yield characteristic glueball sizes in the range $0.1$--$0.3\,\mathrm{fm}$ \cite{Schafer:1994fd, Forkel:2000fd, PhysRevD.43.2301, deForcrand:1991kc, Loan:2006gm}.

\Cref{tab:lambda} shows that the $\Lambda_j$ increases with the vacuum gluon condensate and decreases with the glueball mass, spanning approximately $0.66$--$1.20$ GeV across representative phenomenological, QCD sum-rule, and lattice determinations. These results demonstrate a clear separation of scales within the EFT: the glueball mass uniquely determines the intrinsic $r_{\rm int}$, whereas the gluon condensate fixes the overall interaction scale through $\Lambda_j$. Consequently, the prediction for $r_{\rm int}$ is robust against the phenomenological uncertainties associated with the determination of the gluon condensate and $\Lambda_j$. 

In \cref{tab:lambda1}, we investigate the effect of the first excited glueball state $0^{++}$ on $r_{\rm int}$. Although $\Lambda_0$ and $\Lambda_1$ vary by nearly a factor of three depending on the gluon-condensate input, the resulting $r_{\rm int}^*$ remains essentially unchanged, $r_{\rm int}^*\simeq0.2402~\mathrm{fm}$. This remarkable stability arises from the interplay between the $m_{\phi_j}$ and $\Lambda_j$. Although the heavier excited state favors a smaller $r_{\rm int}$, the correlated dependence of the $\Lambda_j$ on the gluon condensate largely cancels in the interaction-radius ratio, making the predicted $14\%$ reduction a consequence of the excited-state contribution rather than an artifact of the condensate determination.

The results in \cref{tab:self_consistent} show a clear inverse correlation between the dimension-two gluon condensate and the dynamically generated glueball mass. With the dimension-four condensate fixed at $0.0280~\mathrm{GeV}^4$, increasing $\langle g^2A^2\rangle_E$ from $3.03$ to $5.10~\mathrm{GeV}^2$ lowers $m_\phi$ from $1.157$ to $0.891~\mathrm{GeV}$. The corresponding strong coupling increases from $\alpha_s(m_\phi^2)=0.407$ to $0.509$, with $g(m_\phi^2)$ increasing from $2.261$ to $2.530$. Meanwhile, $\langle A^2\rangle_E$, $|D_M^{(2)}|$, and $\Lambda_j$ increase with the dimension-two condensate, indicating that stronger nonperturbative gluonic correlations are associated with a lower characteristic $m_\phi$ and a larger $\Lambda_j$ within the present parametrization. 

It should be noted that \cref{tab:self_consistent} is intended primarily to assess the characteristic magnitude of the dynamically generated glueball mass within the present framework and to provide a controlled test of its predictive behavior. For the subsequent analysis, however, we adopt glueball masses obtained from lattice QCD and phenomenological determinations, together with the dimension-four gluon condensate, as input parameters. The gauge-dependent dimension-two condensate, $g^2\langle A_\mu^a A^{a\mu}\rangle_E$, is therefore used only in the self-consistent mass analysis and does not enter the phenomenological calculations that follow.

\begin{figure}[t!]
   \centering	
	\includegraphics[width=0.5\textwidth]{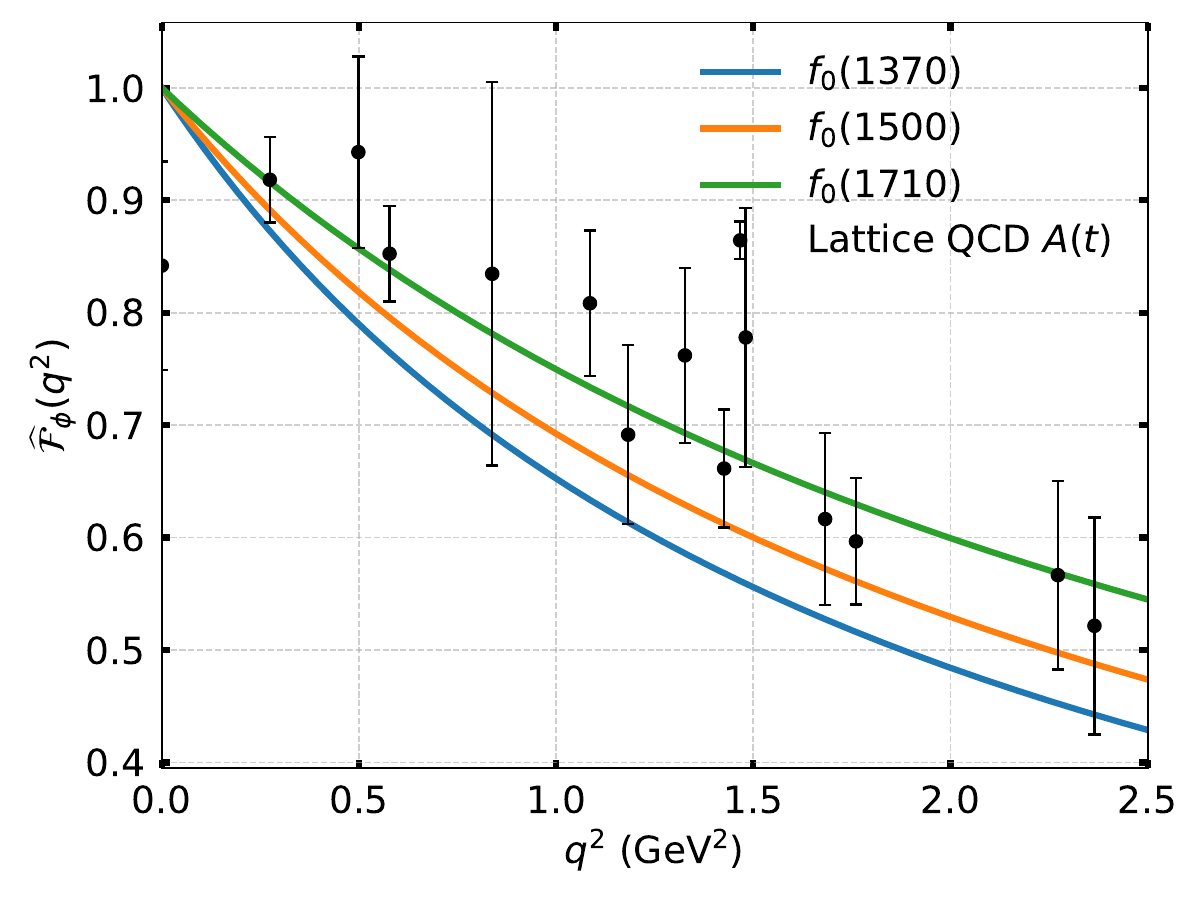}
    	\caption{Normalized scalar glueball form factor, $\widehat{\mathcal{F}}_{\phi}(q^2)$, for the $f_0(1370)$, $f_0(1500)$, and $f_0(1710)$ candidates, shown together with the lattice QCD $A(-t)$ data for visual comparison. For spacelike momentum transfer, $q^2=-t>0$. The glueball mass produces a progressively harder momentum dependence and a smaller effective interaction radius, $r_{\mathrm{int}}=\sqrt{6}/m_\phi$. 
        }
\label{analytic_form}
	\end{figure}
\Cref{analytic_form} illustrates the intrinsic momentum dependence of the normalized scalar glueball form factor for the three candidate ground states. Although all curves are normalized to unity at vanishing momentum transfer, their momentum evolution exhibits a clear dependence on the glueball mass. Heavier scalar glueballs produce a slower falloff with increasing $q^2$, corresponding to a harder form factor and, equivalently, a smaller interaction radius through $r_{\rm int}=\sqrt{6}/m_\phi$. This sensitivity of the form factor to the glueball mass provides a direct means of discriminating among competing scalar glueball candidates. In the following, this qualitative behavior is quantified by fitting the EFT monopole and multipole parameterizations to the lattice Yang--Mills gravitational form factor, thereby identifying the resonance that best reproduces the nonperturbative momentum dependence.

\begin{figure}[t!]
   \centering	
	\includegraphics[width=0.5\textwidth]{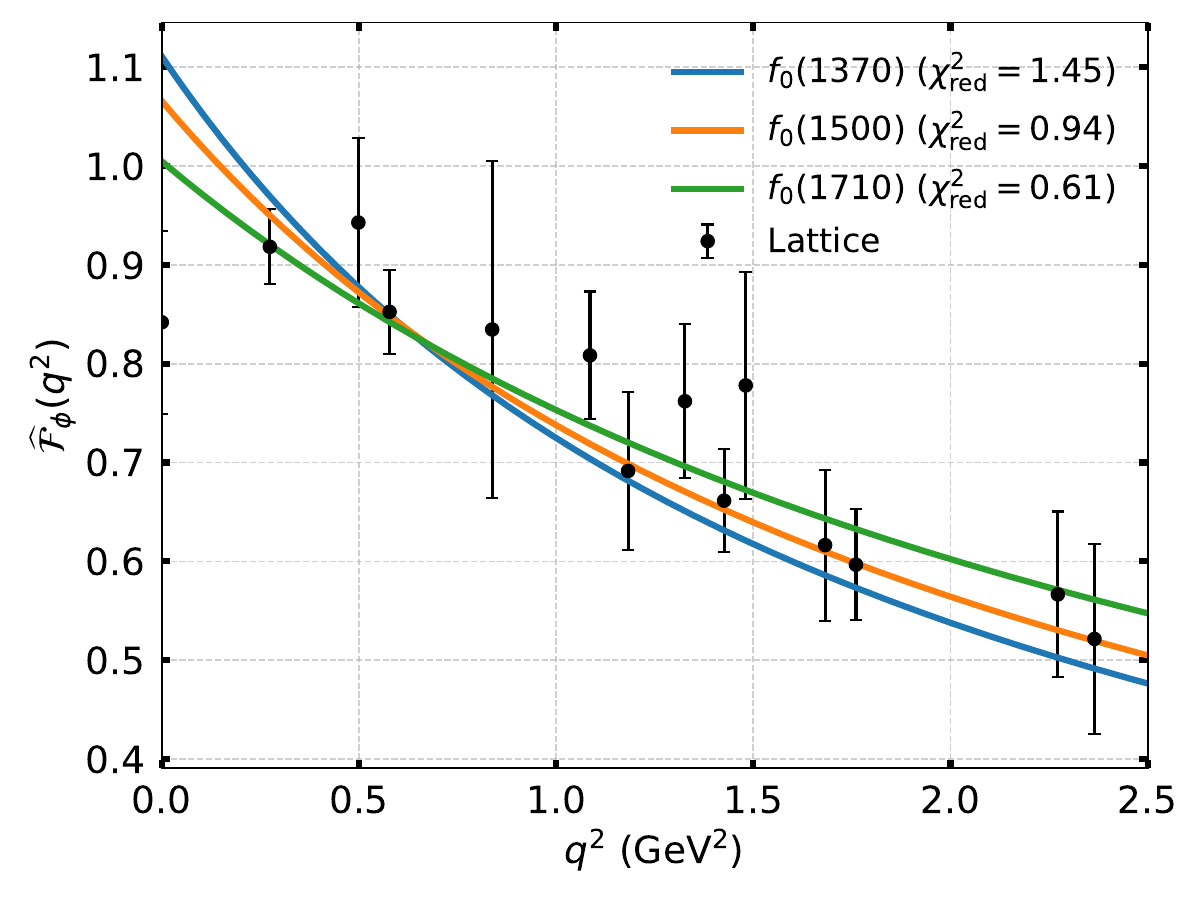}
    	\caption{Comparison between the lattice Yang--Mills scalar glueball gravitational form factor \cite{Abbott:2025irb} and the monopole form factor predicted by the present EFT for the scalar glueball candidates $f_0(1370)$, $f_0(1500)$, and $f_0(1710)$. The curves are obtained from $N\,m_\phi^2/(q^2+m_\phi^2)$, with the glueball masses fixed to their corresponding values. The normalization factor $N$ accounts for the different operator normalizations of the EFT and lattice matrix elements, allowing a comparison of the momentum dependence. The reduced chi-square values, $\chi_{\rm red}^2$, quantify the agreement with the lattice data, favoring the $f_0(1710)$ candidate. 
        }
\label{lattice_comp}
	\end{figure}
\Cref{lattice_comp} shows that the monopole form factors for the three selected glueball candidates reproduce the overall decreasing trend of the lattice form factor with increasing momentum transfer. Their quantitative agreement with the lattice data is assessed through the fits presented in \cref{tab:chi2}. The agreement improves systematically with increasing glueball mass, with the reduced chi-square decreasing from $\chi_{\rm red}^2=1.45$ for $f_0(1370)$ to $0.94$ for $f_0(1500)$ and $0.61$ for $f_0(1710)$. The superior agreement obtained for the $f_0(1710)$ candidate provides independent dynamical support for its interpretation as the dominant scalar glueball and complements the effective interaction-radius analysis. Together, these results demonstrate that the present EFT successfully reproduces both the characteristic spatial scale and the momentum dependence of the scalar glueball, establishing a nontrivial connection between the EFT description and first-principles lattice Yang--Mills calculations. 

In addition to $f_0(1710)$ yielding the lowest $\chi_{\rm red}^2$, from  \cref{tab:chi2}, the fitted normalization factor approaches unity as the assumed glueball mass increases, with $N=1.1109$, $1.0657$, and $1.0047$ for the $f_0(1370)$, $f_0(1500)$, and $f_0(1710)$ candidates, respectively. Since $N$ accounts for the difference between the EFT operator normalization and the lattice gravitational matrix element, the result $N\simeq1$ for the $f_0(1710)$ candidate indicates that the EFT reproduces both the momentum dependence and the overall normalization of the lattice form factor with essentially no additional rescaling, providing independent support for its identification as the lightest scalar glueball. 

\begin{figure}[t!]
   \centering	
	\includegraphics[width=0.5\textwidth]{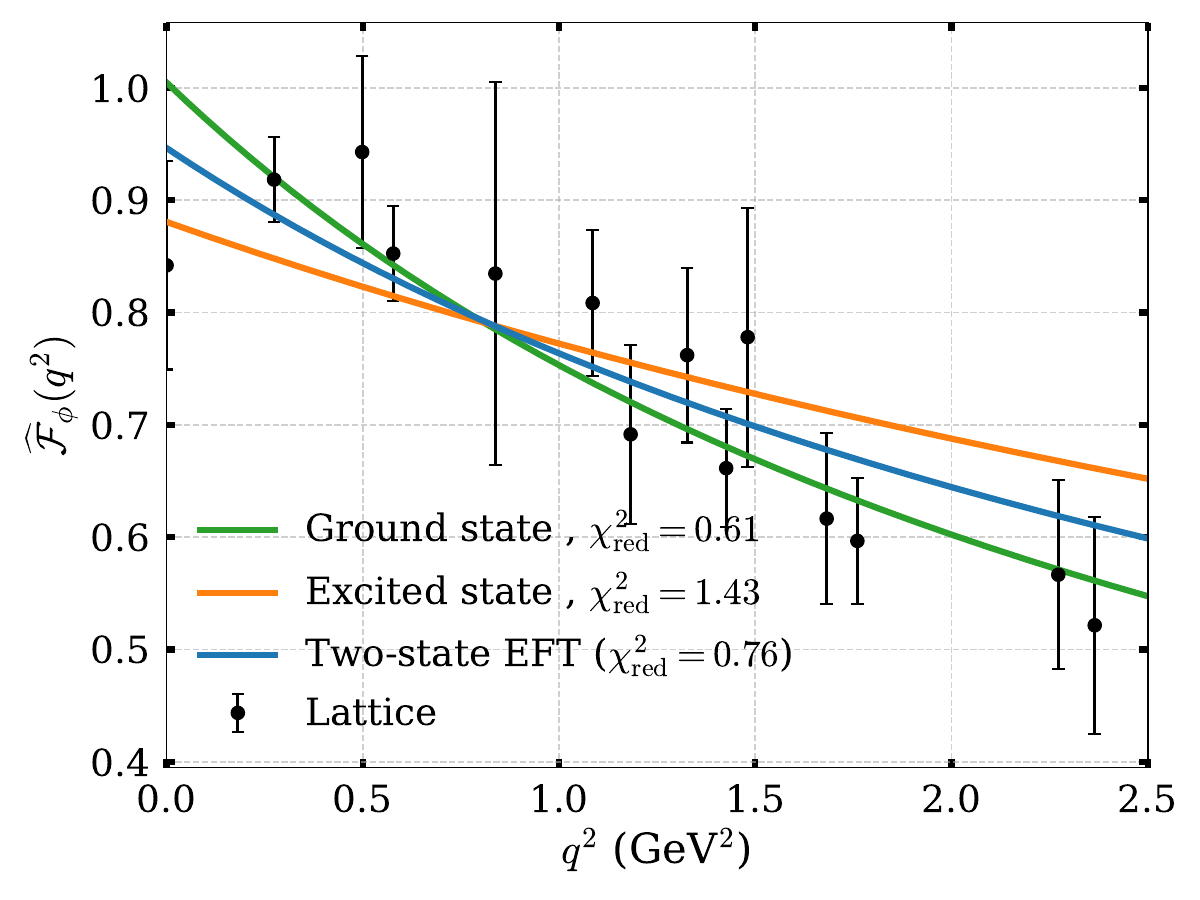}
    	\caption{Scalar glueball form factor $\widehat{\mathcal{F}}_{\phi}(q^2)$ as a function of the momentum transfer $q^2$. The lattice data are compared with single-state monopole fits for the ground-state glueball ($m_{\phi_0}=1.73~\mathrm{GeV}$, $\chi_{\rm red}^2=0.61$) and the first excited scalar glueball ($m_{\phi_1}=2.67~\mathrm{GeV}$, $\chi_{\rm red}^2=1.43$), together with the two-state EFT prediction ($\chi_{\rm red}^2=0.76$), in which the relative contributions of the two glueball states are fixed by the EFT coefficients $c_0$ and $c_1$, while only an overall normalization factor is fitted to the lattice data.}
\label{groud_excited}
	\end{figure}
For the numerical comparison of the multipole expansion, \cref{tawff}, with the lattice gravitational form factor shown in \cref{groud_excited}, we adopt the same fitting procedure as in the monopole analysis, introducing a single overall normalization factor $N$. The quantity fitted to the lattice data is therefore $N\widehat{\mathcal F}_\phi(q^2)$ so that the effective weights entering the fit are $Nc_j$ rather than $c_j$. The coefficients $c_j$ themselves are not free parameters but are uniquely determined by the EFT through the effective interaction scales and glueball masses. For the two-state expansion,
\begin{align}
c_0 &= \frac{\Lambda_1^2 m_{\phi_1}^2}
{\Lambda_0^2 m_{\phi_0}^2+\Lambda_1^2 m_{\phi_1}^2},
\qquad
c_1 = \frac{\Lambda_0^2 m_{\phi_0}^2}
{\Lambda_0^2 m_{\phi_0}^2+\Lambda_1^2 m_{\phi_1}^2}.
\label{cefft}
\end{align}
By construction, $c_0+c_1=1$. Consequently, the fit determines only the single parameter $N$, while the relative contributions of the ground-state and first excited scalar glueballs remain fixed by the EFT.

\Cref{groud_excited} compares the lattice scalar glueball form factor with the single-state monopole and two-state EFT predictions. The ground-state monopole ($m_{\phi_0}=1.73~\mathrm{GeV}$) provides the best individual description of the lattice data, yielding $\chi_{\rm red}^2=0.61$ with $N\approx1.0$, whereas the first excited state ($m_{\phi_1}=2.67~\mathrm{GeV}$) alone gives a significantly poorer fit ($\chi_{\rm red}^2=1.43$, $N=0.88$). The two-state EFT gives an intermediate fit quality, with $\chi^2_{\rm red}=0.76$, improving upon the excited-state-only fit but not the ground-state monopole. These results indicate that the ground-state $0^{++}$ glueball dominates the long-distance behavior of the form factor, whereas the first excited state provides a subleading short-distance correction. This is consistent with \cref{lattice_comp} and \cref{tab:chi2}, where the $f_0(1710)$ candidate yields the lowest reduced chi-square among the single-state fits, while the two-state EFT provides a systematic extension by incorporating excited-state contributions without introducing additional free relative parameters.

 \begin{figure*}[t!]
   \centering	
	\includegraphics[width=1\textwidth]{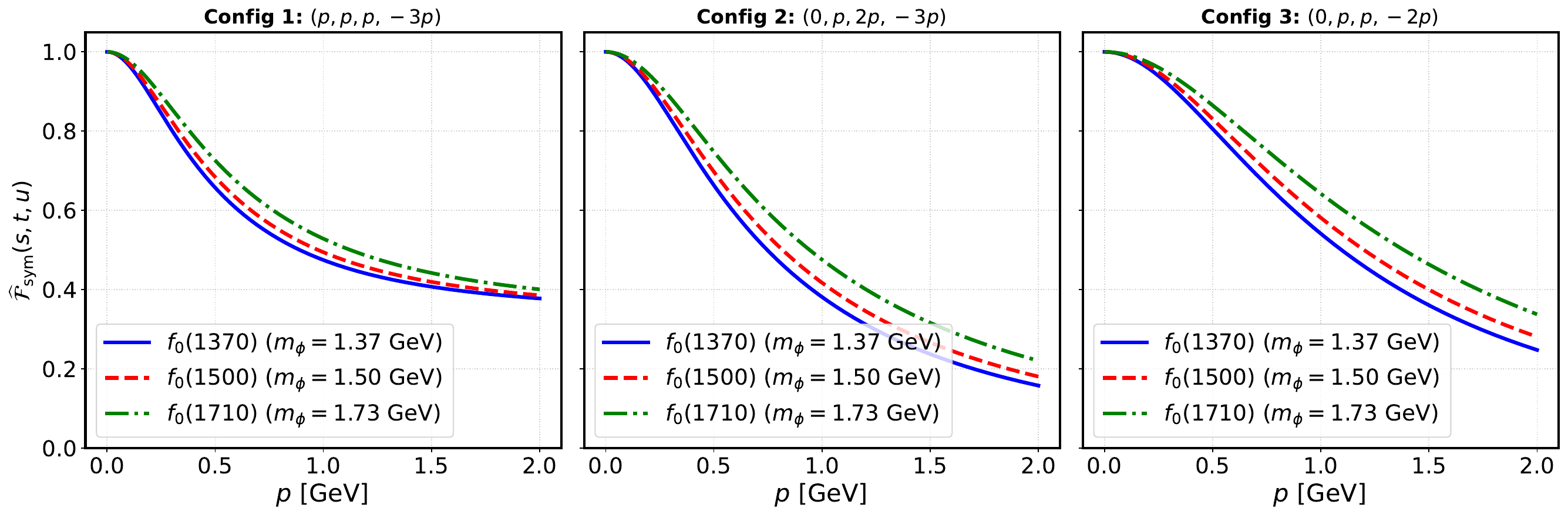}
    	\caption{Mass dependence of the crossing-symmetric, normalized scalar glueball form factor $\widehat{\mathcal{F}}_{\text{sym}}(s,t,u)$ as a function of Euclidean momentum $p$ for the $f_0(1370)$, $f_0(1500)$, and $f_0(1710)$ states across representative kinematic configurations. We summarize the corresponding kinematic values sampled by the $s$, $t$, and $u$-channels in \cref{tab:kinematics}.} 
\label{formfac}
	\end{figure*}
The normalized crossing-symmetric scalar glueball form factor, $\widehat{\mathcal F}_{\rm sym}(s,t,u)$, shown in \cref{formfac} provides a natural quantity for comparison with the lattice studies of the four-gluon vertex by \cite{Colaco:2024gmt, Oliveira:2025tfq}. In their tensor decomposition, the color-singlet tensor $\widetilde{\Gamma}^{(2)}$ is the closest lattice counterpart to the scalar $J^{PC}=0^{++}$ projection considered here, as both probe the same nonperturbative scalar color-singlet channel. However, whereas the lattice calculations determine tensor dressing functions of the full four-gluon vertex, the present EFT isolates the scalar glueball contribution by projecting onto the $J^{PC}=0^{++}$ channel and factorizing the projected amplitudes into universal Lorentz--color kinematic structures and an intrinsic scalar glueball function. The resulting crossing-symmetric normalized form factor therefore encodes the scalar-glueball momentum dependence after combining the contributions from the three Mandelstam channels, while retaining their distinct kinematic arguments. By adopting the one-dimensional momentum configurations used in lattice simulations, the present framework permits a direct qualitative comparison with existing lattice results. The normalized crossing-symmetric form factor also provides a reference for future lattice studies of the scalar $0^{++}$ component of the four-gluon vertex and for phenomenological descriptions of glueball dynamics.

\begin{figure*}[t!]
   \centering	
	\includegraphics[width=1\textwidth]{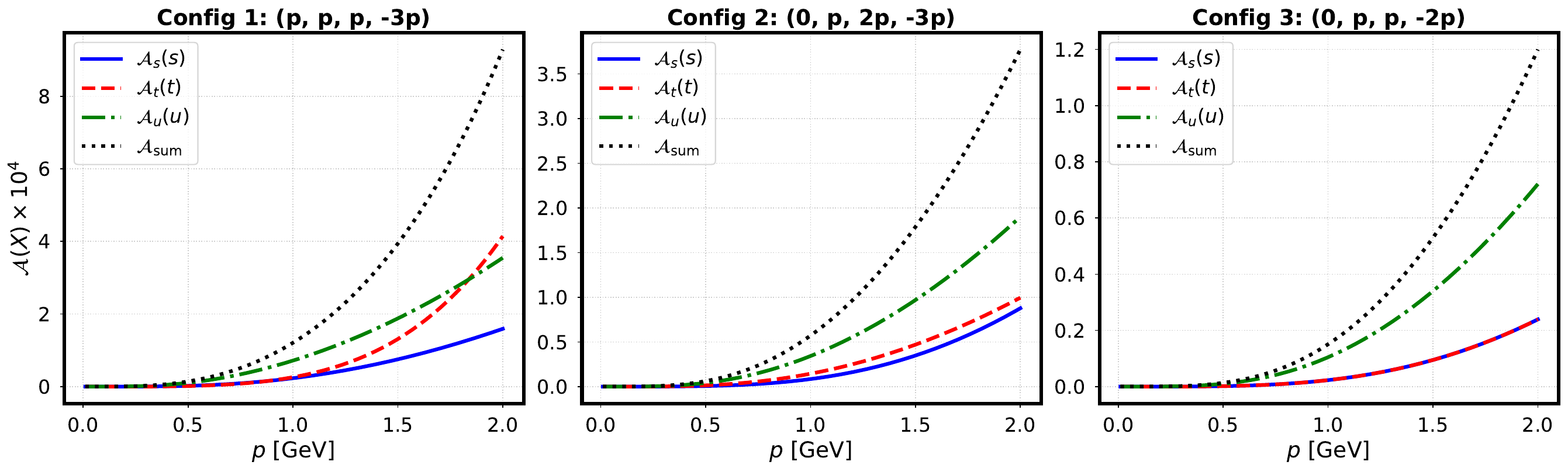}
    	\caption{Scalar projections of the four-gluon scattering amplitude, $\mathcal{A}(s,t,u)$, for the ground-state ($m_{\phi_0}=1.73$ GeV) and first excited ($m_{\phi_1}=2.67$ GeV) scalar glueballs in three representative kinematic configurations. The colored curves correspond to the individual $s$, $t$, and $u$-channel contributions, while the black dotted curve denotes the total crossing-symmetric amplitude, $\mathcal{A}_{\rm sum}=\mathcal{A}_s+\mathcal{A}_t+\mathcal{A}_u$. }
\label{ampltude1}
	\end{figure*}

\Cref{ampltude1} shows that the qualitative structure of the four-gluon scattering amplitude is largely independent of the chosen momentum configuration. Although the relative magnitudes of the individual $s$-, $t$-, and $u$-channel contributions vary with the external kinematics, their coherent sum exhibits a smooth and systematic momentum dependence in all three cases. The dominant channel changes with the momentum configuration, reflecting the redistribution of scattering strength among the Mandelstam channels while preserving the overall crossing-symmetric structure of the amplitude.

\section{Conclusions}\label{conc}

This work establishes a quantitative EFT for scalar glueball interactions rooted in the nonperturbative Yang--Mills vacuum. A central result is that the interaction radius (see \cref{eq:rint}) of the ground-state scalar glueball emerges directly from the pole structure of the normalized form factor and is completely determined by the glueball mass, independent of both the effective interaction scale and the vacuum gluon condensate. For the $f_0(1710)$ candidate, the predicted interaction radius is in close agreement with recent lattice Yang--Mills determinations \cite{Abbott:2025irb}, while the normalized form factor reproduces the lattice gravitational form factor with the lowest reduced chi-square. Furthermore, the fitted normalization factor is consistent with unity, indicating that only a small rescaling is required to match the normalization of the lattice matrix element while preserving its momentum dependence. Taken together with the agreement in the interaction radius, these results provide quantitative support for the $f_0(1710)$ assignment as the scalar glueball candidate favored within the present analysis, while indicating that its long-distance structure is dominated by the ground-state pole.

Extending the formalism to a two-state description shows that excited scalar glueballs contribute primarily through short-range corrections. Although the first excited $0^{++}$ state reduces the interaction radius by approximately $14\%$, the resulting prediction remains compatible with the lattice determination within $\sim0.7\sigma$ and is remarkably insensitive to the gluon-condensate input, varying by less than $0.03\%$ across representative lattice and QCD sum-rule determinations. More generally, the multipole expansion provides a systematic way to incorporate higher scalar-glueball excitations. The one-variable form factor retains its low-energy normalization, while crossing symmetry is recovered through its symmetric evaluation in the $s$, $t$, and $u$ channels. The gluon condensate determines $\Lambda_j$, whereas $m_{\phi_j}$ and their associated weights govern the momentum dependence and spatial structure of the normalized form factor. Overall, the framework establishes a systematic connection between the EFT description and lattice Yang--Mills calculations. It provides quantitative benchmarks for future studies of scalar-glueball interaction radii, gravitational form factors, and the $0^{++}$ component of the four-gluon vertex, and can be extended to include higher-order corrections, additional glueball excitations, and unquenched QCD.

Beyond the interaction-radius analysis, the present framework introduces a normalized one-variable scalar-glueball form factor, $\widehat{\mathcal F}_{\phi}(q^2)$, whose argument is evaluated independently in the $s$, $t$, and $u$ channels. A crossing-symmetric combination is then obtained by averaging the three channel contributions, providing a useful quantity for comparison with lattice studies of the four-gluon vertex \cite{Colaco:2024gmt, Oliveira:2025tfq}. For the lattice-motivated kinematic configurations considered here, the resulting form factor exhibits a smooth momentum dependence. Including the first excited $0^{++}$ state produces a moderate modification at intermediate and larger momentum transfers while preserving the normalization at vanishing momentum transfer. The scalar-projected four-gluon amplitudes retain the distinct channel dependence arising from the external kinematics and decrease systematically with increasing glueball mass. These results provide quantitative benchmarks for future lattice studies of the scalar color-singlet component of the four-gluon vertex and for investigating the contribution of excited scalar glueball states.

We emphasize that the exponential dressing of the gauge field is a physically motivated effective ansatz inspired by the coherent-state structure of the nonperturbative Yang--Mills vacuum, rather than a first-principles derivation from the fundamental Yang--Mills theory. The effective Lagrangian is constructed consistently through $\mathcal{O}(\phi^2)$, while the glueball masses are taken as phenomenological or lattice inputs when determining $\Lambda_j$. In addition, the nonperturbative condensate analysis provides a self-consistent determination of the scalar glueball mass within the adopted EFT framework, subject to the dimension-two and dimension-four condensate inputs. The interaction radius, $r_{\rm int}={\sqrt{6}}/{m_\phi}$, nevertheless remains a particularly robust observable of the framework, directly relating the glueball mass scale to the spatial extent of the effective interaction and providing a useful benchmark for comparison with lattice determinations.

\begin{acknowledgments}
A.~I. and T.~F.  thank CFisUC for hosting the collaboration.
A. I. acknowledges financial support from the São Paulo State Research Foundation (FAPESP), Grant Nos. 2023/09545-1 and 2025/17347-0. 
T. F. thanks to FAPESP, Grant Nos. 2023/13749-1, 2024/17816-8 and 2025/05312-8, and to the Conselho Nacional de Desenvolvimento 
Cient\'{i}fico e Tecnol\'{o}gico (CNPq), Grant No. 306834/2022-7. This work is a part of the
project Instituto Nacional de  Ci\^{e}ncia e Tecnologia - F\'{\i}sica
Nuclear e Aplica\c{c}\~{o}es  Proc. No. CNPq 408419/2024-5. 
This work was partially supported by national funds from FCT (Fundação para a Ciência e a Tecnologia, I.P, Portugal) under project UID/04564/2025, identified by DOI 10.54499/UIDB/04564/2025.

\end{acknowledgments}

\bibliography{refrences}

\end{document}